\documentclass[11pt,letterpaper]{article}
\usepackage{jheppub}

\usepackage{graphicx}
\usepackage{bbm}
\usepackage{amsmath}
\usepackage{amssymb}
\usepackage{mathtools}
\usepackage{amsthm}
\usepackage{mathrsfs}
\usepackage{marvosym}
\usepackage{dsfont}
\usepackage[labelformat=simple]{subcaption}
\usepackage{xcolor}
\usepackage{braket}
\usepackage{cleveref}
\usepackage{comment}
\usepackage{float}
\usepackage{fancybox}
\usepackage{slashed}
\usepackage[skins,theorems]{tcolorbox}
\tcbset{highlight math style={enhanced,
		colframe=black,colback=white,arc=0pt,boxrule=1pt}}
\usepackage{overpic}

\allowdisplaybreaks

\definecolor{dark-gray}{gray}{0.20}
\definecolor{gray}{gray}{0.30}
\definecolor{light-gray}{gray}{0.80}
\definecolor{dark-red}{rgb}{0.7,0,0}
\definecolor{dark-green}{rgb}{0.1,0.4,0}
\definecolor{dark-blue}{rgb}{0.3,0.3,0.7}
\definecolor{light-blue}{rgb}{0.8,0.8,1}
\definecolor{swamp}{RGB}{240, 199, 197}

\newcommand{\be}{\begin{equation}}
	\newcommand{\ee}{\end{equation}}

\def\be{\begin{equation}}
	\def\ee{\end{equation}}
\def\bea{\begin{eqnarray}}
	\def\eea{\end{eqnarray}}

\newcommand{\beq}{\begin{equation}}  \newcommand{\eeq}{\end{equation}}
\newcommand{\bal}{\begin{aligned}}   \newcommand{\eal}{\end{aligned}}
\def\beqa{\begin{eqnarray}}
	\def\eeqa{\end{eqnarray}}

\DeclareMathOperator{\csch}{csch}

\numberwithin{equation}{section}

\def\simleq{\; \raise0.3ex\hbox{$<$\kern-0.75em
		\raise-1.1ex\hbox{$\sim$}}\; }
\def\simgeq{\; \raise0.3ex\hbox{$>$\kern-0.75em
		\raise-1.1ex\hbox{$\sim$}}\; }

\numberwithin{equation}{section}

\hypersetup{
	colorlinks=true,
	linkcolor=dark-blue,
	citecolor=dark-red,
	urlcolor=dark-green,
	linktoc=page
}

\theoremstyle{remark}

\newtheoremstyle{named}{}{}{\itshape}{}{\bfseries}{.}{.5em}{#3}
\theoremstyle{named}

\title{\centering  On Supersymmetric D-brane Probes\\ in 4d $\mathcal{N}=2$ $\text{AdS}_2\times \mathbf{S}^2$ Attractors}

\author{Alberto Castellano$^{1,2}$,} 
\author{Carmine Montella$^{3}$,}
\author{Matteo Zatti$^{3}$}
\affiliation{$^1$Enrico Fermi Institute \& Leinweber Institute for Theoretical Physics,\\
University of Chicago, Chicago, IL 60637, USA}
\affiliation{$^2$Kavli Institute for Cosmological Physics,\\
University of Chicago, Chicago, IL
60637, USA}
\affiliation{$^3$Max-Planck-Institut f\"ur Physik (Werner-Heisenberg-Institut),\\
Boltzmannstrasse 8, 85748 Garching bei M\"unchen, Germany}

\emailAdd{acastellano@uchicago.edu, montella@mpp.mpg.de, zatti@mpp.mpg.de}

\abstract{We extend the $\kappa$-symmetry analysis of supersymmetric D-brane probes in the $\mathrm{AdS}_2 \times \mathbf{S}^2$ attractor geometry, originally performed by Simons, Strominger, Thompson, and Yin, to also include stationary---but non-static---worldlines carrying angular momentum along the 2-sphere. We demonstrate that certain special trajectories, with fixed radius and orbital velocity, solve the equations of motion and moreover satisfy a supersymmetry preserving condition, thus defining new $\frac12$-BPS configurations. Furthermore, these classical paths are shown to saturate a lower bound for the Hamiltonian generating global time translations, with the corresponding minimal energy depending on a generalized angular momentum vector $\boldsymbol{J}$. The direction of the latter, in turn, determines exactly which supercharges remain unbroken. Our results reveal a richer spectrum of (multi-particle) supersymmetric states in $\mathrm{AdS}_2 \times \mathbf{S}^2$, which can be organized into distinct selection sectors labeled by the conserved $SU(2)$ charges. This construction has direct applications in black hole microstate counting, the analysis of probe dynamics, and $\text{AdS}_2/\text{CFT}_1$ holography.}

\begin{document}

	\makeatletter
	\let\old@fpheader\@fpheader
	\renewcommand{\@fpheader}{\vspace*{-0.1cm} \hfill EFI-25-10\\ \vspace* {-0.1cm} \hfill MPP-2025-148}
	\makeatother
	\maketitle
	\setcounter{page}{1}
	\pagenumbering{roman}

\newcommand{\remove}[1]{\textcolor{red}{\sout{#1}}}

\newpage
\pagenumbering{arabic} 

\section{Introduction and Discussion}
\label{s:intro}

The study of black hole solutions in string theory provides one of the most promising windows into the quantum structure of space and time. In this regard, supersymmetric black holes play a particularly distinguished role. Not only are they under considerable computational control thanks to supersymmetry, but they also allow for a precise accounting of the microscopic black hole entropy within a consistent theory of quantum gravity \cite{Strominger:1996sh, Maldacena:1997de, Vafa:1997gr}. A prototypical setting where this analysis becomes tractable arises in Type IIA string theory compactified on Calabi–Yau threefolds. In such setups, BPS black holes sourced by D-brane bound states give rise to near-horizon geometries of the form $\mathrm{AdS}_2 \times \mathbf{S}^2 \times \mathrm{CY}_3$, with fluxes fixed and moduli stabilized by the attractor mechanism in terms of the underlying electromagnetic charges \cite{Ferrara:1995ih,Strominger:1996kf,Ferrara:1996dd,Ferrara:1996um}.

A natural way to extract information from these backgrounds is to introduce a D-brane probe into the AdS$_2 \times \mathbf{S}^2$ throat and study its target-space dynamics. The identification of supersymmetric trajectories is furthermore of particular significance. On one hand, they correspond to stable configurations that preserve a fraction of the ambient supersymmetry, thereby enabling explicit analytical computations. On the other hand, these trajectories may yield valuable insights into the quantum stability of the solution and even constitute the relevant saddles of the Euclidean worldline path integral. Relatedly, this framework can be employed to compute (non-)perturbative corrections to black hole thermodynamics \cite{Maldacena:1998uz, Pioline:2005pf, Castellano:2025yur}, and perhaps to formulate a holographic description of the underlying system \cite{Strominger:1998yg,Gaiotto:2004ij,Gaiotto:2004pc, Denef:2007yt, Azeyanagi:2007bj,Sen:2008vm}.

These supersymmetry-preserving configurations can be studied via the $\kappa$‑symmetric worldline action, which encodes the couplings of the probe to the background fluxes and curvature \cite{Becker:1995kb, Bergshoeff:1997kr, Billo:1999ip, Simon:2011rw}. In \cite{Simons:2004nm}, it was moreover shown that \textit{static} BPS solutions do in fact exist. Their radial position in $\mathrm{AdS}_2$ is determined entirely by the phase of the central charge $Z$ associated to the D-brane gauge charges. Remarkably, it was found that not only single-particle but also multi-particle states---each one carrying a priori different central charges $Z_i$---can preserve a common set of supersymmetries. Unlike asymptotically flat compactifications, where the same condition requires the charge vectors to be mutually aligned, the enhanced superconformal symmetry of the near-horizon $\mathrm{AdS}_2 \times \mathbf{S}^2$ geometry allows in general misaligned multi-centered configurations to break identical supercharges. This property reflects the more intricate structure of supersymmetric bound states supported by the attractor point. Further studies on the topic were pursued in \cite{Li:2006uq}, where an analogous problem was addressed in the context of five-dimensional BMPV black holes \cite{Breckenridge:1996is}. There, due to the nature of the rotating solution, the authors identified $\frac{1}{2}$-BPS motions corresponding to \emph{stationary} D-brane probes orbiting along angular directions of the internal (squashed) $\mathbf{S}^3$. 

\medskip

This naturally leads to the following  important question: Do stationary, orbiting BPS particle configurations exist in the four-dimensional $\mathrm{AdS}_2 \times \mathbf{S}^2$ attractor geometry? In other words, can one construct supersymmetric worldlines that carry angular momentum along the sphere, while remaining localized in the radial direction of the 2d Anti-de Sitter component?

\clearpage

Such configurations would give additional contributions to the indexed partition function of the worldline theory arising from sectors with non-minimal $SU(2)$ angular momentum. Their existence would then complete the parallel with the analysis performed in \cite{Simons:2004nm,Li:2006uq}, as well as improve our understanding of the space of supersymmetric excitations in four-dimensional attractor backgrounds.\footnote{\label{fnote:AdS3CFT2}The existence of these configurations was implicitly assumed by \cite{Gaiotto:2006ns} in an attempt to derive the OSV formula \cite{Ooguri:2004zv} through direct computation of the elliptic genus for a particular class of 4d BPS black holes using duality with 5d M-theory and AdS$_3$/CFT$_2$. This required considering chiral primary states with all possible R-charges in the conformal theory, corresponding to different generalized angular momenta on $\mathbf{S}^2$.} Additionally, the inclusion of orbiting BPS probes may be essential to capturing both perturbative and non-perturbative D-brane instanton corrections to protected quantities such as supersymmetric indices and quantum entropy functions \cite{Sen:2007qy,Dabholkar:2010uh,Dedushenko:2014nya, Iliesiu:2022kny,Cassani:2025sim, Castellano:2025yur}, which in turn are central to the microscopic interpretation of black hole entropy and the OSV conjecture \cite{Ooguri:2004zv}. For these reasons, the presence of angular momentum degrees of freedom---often overlooked in the literature---underscores the significance of such contributions within the gravitational framework of extremal black holes. By bridging the microscopic D-brane constituents with macroscopic gravitational observables, our results might offer valuable clues as to how brane interactions could be holographically encoded within the dual quantum mechanical system of the AdS$_2$/CFT$_1$ correspondence \cite{Strominger:1998yg, Sen:2008vm, Azeyanagi:2007bj}. We hope that this work will provide both useful tools and interesting insights that could contribute toward a deeper understanding of non-perturbative phenomena in black hole physics within string theory.

\subsection{Summary of results}\label{ss:summary}

In this note, we refine and extend the analysis of supersymmetric particle probes in the four-dimensional $\mathcal{N}= 2$ $\mathrm{AdS}_2 \times \mathbf{S}^2$ attractor geometry by including stationary configurations carrying non-vanishing angular momentum ($\ell$) along the sphere. Building on the works \cite{Simons:2004nm, Li:2006uq}, we identify a new family of BPS worldlines at constant radius in Anti-de Sitter and fixed polar angle
\begin{equation}\label{eq:susystationary}
\sinh \chi = \frac{q_e}{|j|}\,,\qquad \cos \theta=-\frac{q_m}{j}\, ,\qquad \frac{d\phi}{d\tau}=\pm1\, ,
\end{equation}
where $(q_e, q_m)$ denote, respectively, the effective electric/magnetic charges of the particle and 
\begin{equation}
 j = \pm \sqrt{q_m^2 + \ell^2} \,,
\end{equation}
represents their generalized angular momentum. These trajectories solve the equations of motion and supersymmetry constraints, thereby saturating a lower bound for the (global) Hamiltonian that depends on the Casimir invariant along $\mathbf{S}^2$. The above result \eqref{eq:susystationary} moreover incorporates the solution already discussed in \cite{Simons:2004nm}, corresponding to a probe with vanishing angular momentum $\ell = 0$. In that case,
\begin{equation}
\tanh \chi = \frac{q_e}{\sqrt{q_m^2 + q_e^2}} = \frac{\text{Re}\,( \bar{Z}_{\rm BH}Z)}{|\bar{Z}_{\rm BH}Z|}\,,\qquad \cos \theta=-\text{sgn}(j/q_m)\, ,
\end{equation}
which describes a static (anti-)particle located at the north/south pole of the 2-sphere.

\clearpage

\begin{figure}[h]
    \centering
\subcaptionbox{\label{sfig:static}}{
		\includegraphics[scale=0.4]{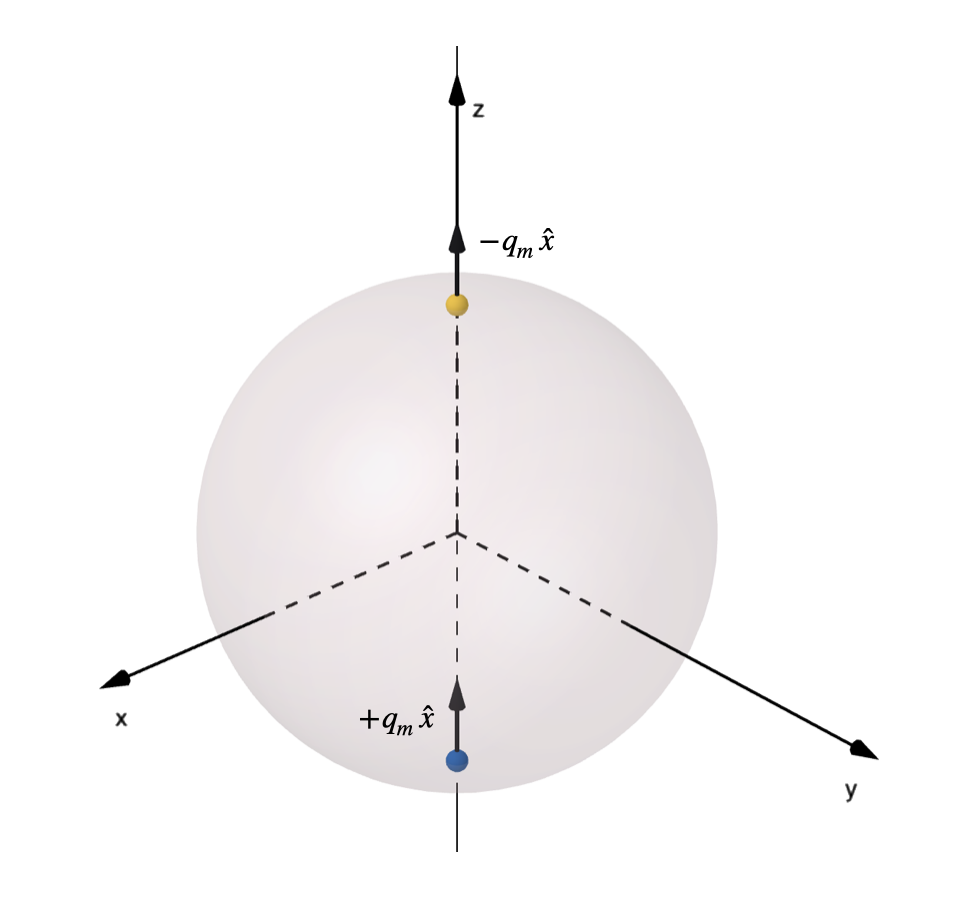}}
        \hspace{0.6cm}
        \subcaptionbox{\label{sfig:stationary}}{\includegraphics[scale=0.35]{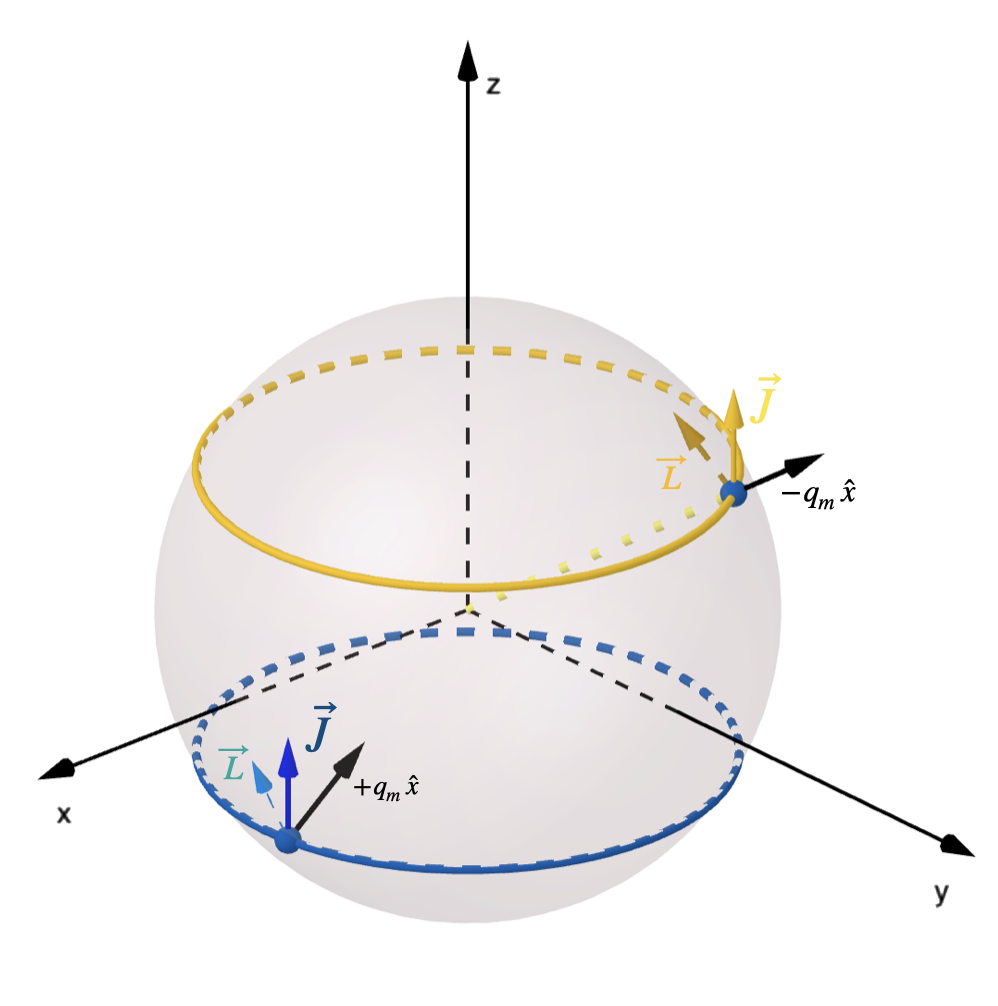}}
		\caption{\small A system comprised by a particle/anti-particle pair can be BPS if the total generalized angular momentum satisfies $|\boldsymbol{J}_{\rm tot}| = |\boldsymbol{J}_1| + |\boldsymbol{J}_2|$. $\textbf{(a)}$ Static configuration with the probes located at antipodal points on $\mathbf{S}^2$. $\textbf{(b)}$ Stationary case with the particles rotating in opposite directions.} 
		\label{fig:multiparticleBPS}
\end{figure}

In addition, we argue that the direction of the generalized angular momentum vector $\boldsymbol{J}$ (cf. eq.~\eqref{eq:SU2generators} for a precise definition) specifies the subset of unbroken supercharges, thus allowing for a richer spectrum of multi-particle BPS configurations where all the individual constituents satisfy \eqref{eq:susystationary}, with their corresponding angular momenta being perfectly aligned, see Figure \ref{fig:multiparticleBPS}. This includes, in particular, situations where both particles and anti-particles are present and remain stationary, as opposed to what happens in 4d Minkowski space.

\medskip

The paper is organized as follows. In Section \ref{s:review}, we present the relevant BPS black hole solutions in 4d $\mathcal{N}=2$ supergravity and study charged geodesics\footnote{By a slight abuse of notation, we oftentimes refer in this work to the classical trajectories that solve the equations of motion---including both gravitational and gauge interactions---as \emph{charged geodesics}, even though, strictly speaking, they do not satisfy the geodesic equation.} probing their near-horizon geometry. We then classify all possible classical trajectories, including those with orbital motion on $\mathbf{S}^2$, highlighting how the enhanced AdS$_2 \times \mathbf{S}^2$ isometries constrain their dynamics. Section \ref{s:staticprobes} focuses on static probe configurations. After reviewing worldline supersymmetry and $\kappa$-symmetry, followed by an explicit construction of the background Killing spinors \cite{Alonso-Alberca:2002wsh}, we rederive the BPS conditions for this kind of systems following the original analysis in \cite{Simons:2004nm}. We also take the opportunity to emphasize certain features that will guide our subsequent generalization. In Section \ref{s:stationaryprobes}, we show that a wider class of non-static---yet stationary---geodesics can similarly preserve half of the spacetime superconformal charges. We derive the corresponding generalized BPS constraints and explain how the unbroken supersymmetries are precisely determined by the direction of the conserved angular momentum.

\clearpage

\section{$\mathrm{AdS}_2 \times \mathbf{S}^2$ Geometry and Charged Particle Dynamics}\label{s:review}

The aim of this section is to introduce the relevant material for the discussions in the upcoming chapters. Thus, we first review the explicit metric and gauge backgrounds of maximally supersymmetric $\mathrm{AdS}_2 \times \mathbf{S}^2$ solutions, obtained by taking the near-horizon limit of BPS black hole geometries in the underlying four-dimensional $\mathcal{N}=2$ supergravity theory. Subsequently, in Section \ref{ss:geodesics} we analyze all classically allowed configurations of charged probe particles in this symmetric spacetime. For a more detailed treatment, we refer the reader to e.g., \cite{Castellano:2025yur}.

\subsection{The metric and gauge backgrounds}\label{ss:background}

\begin{figure}[t!]
	\begin{center}
		\centering
		\includegraphics[width=0.4\linewidth]{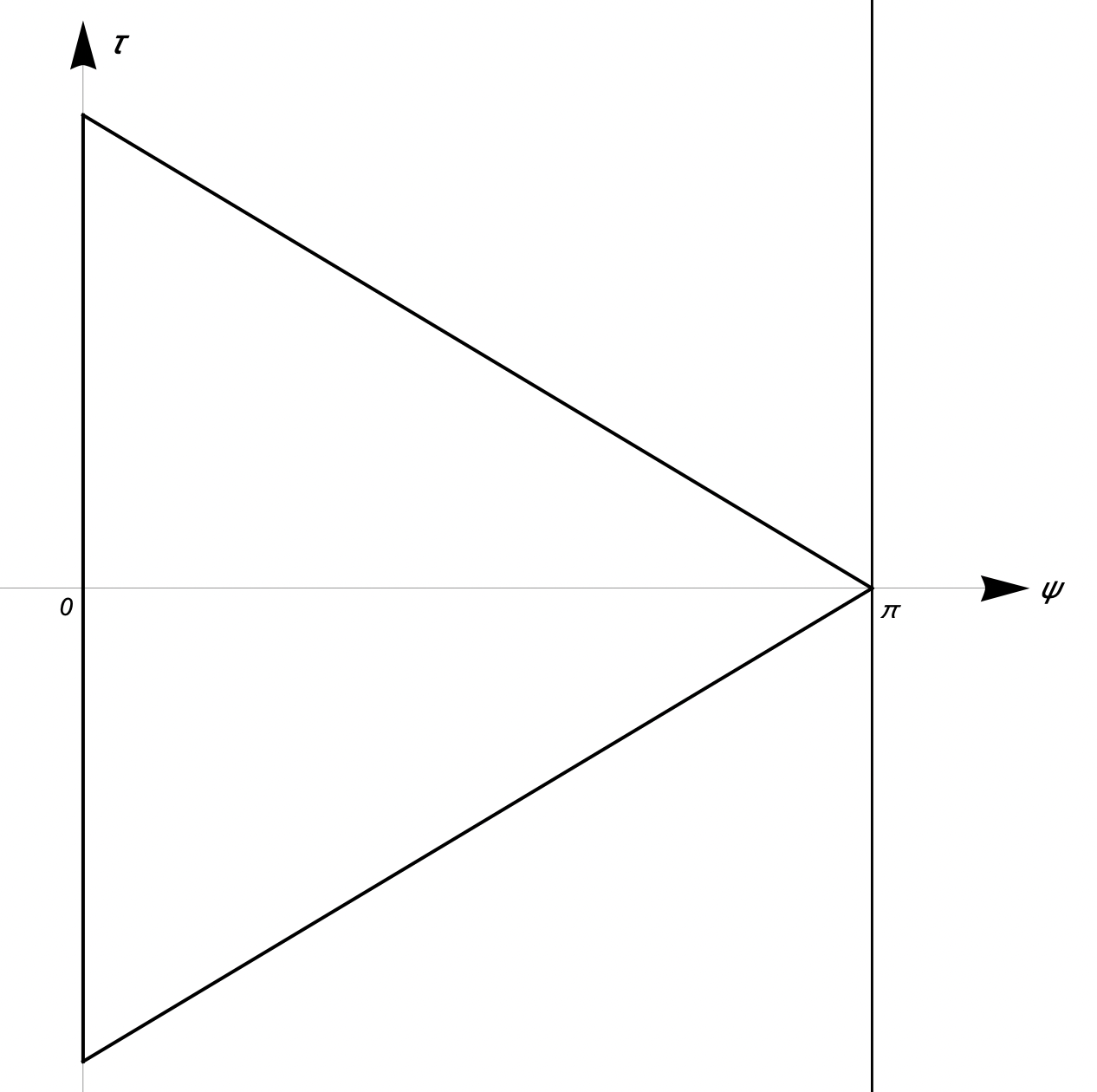}
        \hspace{0.6cm}
		\caption{\small Penrose diagram of 2d Anti-de Sitter space in Poincaré and global (strip) coordinates. The triangular region corresponds to a single Poincaré patch, whereas global AdS contains an infinite sequence of such consecutive slices.} 
		\label{fig:PenroseAdS}
	\end{center}
\end{figure}

As is familiar from our experience with string theory and holography, a useful procedure to obtain supersymmetric $\mathrm{AdS}_p \times \mathbf{S}^q$ flux vacua consists in placing a stack of D-branes and considering the near-horizon limit of their backreacted geometry \cite{Gibbons:1993sv,Maldacena:1997re,Gubser:1998bc,Witten:1998qj,Aharony:1999ti}. In the present case, the 4d solitonic objects sourcing the solution are BPS black holes \cite{Strominger:1998yg,Maldacena:1998uz}, whose line element close to the horizon---located at $r=r_h$---has the form 
\begin{equation}\label{eq:BertRobmetric}
ds^{2} = -\frac{y^{2}}{r_h^{2}} dt^{2} + \frac{r_h^{2}}{y^{2}} dy^{2} + r_h^{2} d\Omega_{2}^{2}\,,
\end{equation}
where we defined a shifted radial coordinate $y = r - r_h$ and we have focused on the region $y/r_h \ll 1$. This corresponds to the Bertotti-Robinson spacetime \cite{Gibbons:1982fy,Gibbons:1987ps,Garfinkle:1990qj}, which describes a conformally flat universe with $\mathrm{AdS}_2 \times \mathbf{S}^2$ topology, as one can make manifest by introducing a new radial coordinate $\rho = r_h^2 / y$, such that \eqref{eq:BertRobmetric} becomes now\footnote{Notice that this change of coordinates reverses the orientation of AdS$_2$, since spacelike infinity (originally situated along $y \to \infty$) appears now at $\rho = 0$, whereas the black hole horizon ($y=0$) corresponds to $\rho \to \infty$. This chart, however, fails to account for the entire 4d spacetime when including the region $y \geq 0$, cf. eq.~\eqref{eq:globalmetricAdS2}.}
\begin{equation}\label{eq:conformalcoords}
ds^{2} = \frac{r_h^{2}}{\rho^2} \left(- dt^{2} + d\rho^2 +\rho^2 d\Omega_{2}^{2} \right)\,.
\end{equation}
More precisely, the above metric covers a single Poincaré patch of AdS$_2$. Therefore, to recover the global structure one may proceed via the usual hypersurface embedding in $\mathbb{R}^{2,1}$ as follows. One starts by introducing global coordinates $(X^0, X^2, X^1)$, and defines an hyperboloid as the solution to the constraint equation
\begin{equation}
    -(X^0)^2 - (X^2)^2 + (X^1)^2 = -r_h^2 := -R^2\, .
\end{equation} 
A convenient parametrization of this surface is given by
\begin{equation}
    X^0 = R \cosh\chi\,\sin\tau\, , \qquad X^2 = R \cosh\chi\,\cos\tau\, , \qquad X^1 = R \sinh\chi\, ,
\end{equation} 
from which the pull-back metric may be recast into the global form
\begin{equation}\label{eq:globalmetricAdS2}
ds^{2} = R^2 \left( - \cosh^2\chi\, d\tau^{2} + d\chi^2 + d\Omega_{2}^{2}\right)\, ,
\end{equation}
where the universal cover of the compact time direction $\tau$ is to be understood. An important feature of two-dimensional Anti-de Sitter space is that the condition $X^1 \in \mathbb{R}$ implies the existence of two disconnected timelike boundaries, which are located at $\chi \to \pm \infty$. Furthermore, starting from the global coordinate system discussed above, one can introduce an alternative parametrization given by
\begin{equation}\label{eq:stripcoords}
\sin{\psi} =\frac{1}{\cosh \chi}\, ,\qquad \text{with}\quad \psi \in [ 0, \pi]\, ,
\end{equation}
which leads to a new representation of the metric tensor
\begin{equation}
ds^2=\frac{R^2}{\sin^2\psi} \left( -d\tau^2+d\psi^2+ \sin^2\psi\, d\Omega_2^2\right) \, ,
\end{equation}
In this chart, the radial coordinate is effectively compactified such that the two conformal boundaries are brought to $\psi = 0$ and $\psi = \pi$, respectively (see Figure~\ref{fig:PenroseAdS}).

\subsection{Classical D-brane charged geodesics}\label{ss:geodesics}

Let us study now the classical trajectories followed by charged BPS particles in the previous AdS$_2\times \mathbf{S}^2$ background. For later convenience, we will parametrize these solutions in terms of the underlying black hole data, following the near-horizon prescription outlined before. Notably, the radius of the AdS$_2$ factor is determined by the central charge of the extremal black hole via the relation
\begin{equation}\label{eq:BHcentralcharge}
R^2=r_h^2= |C|^2 e^{-K}=|Z_{\rm BH}|^2\, ,\qquad \text{with}\quad C=e^{K/2}\bar{Z}_{\rm BH}\, ,
\end{equation}
where $K$ is the  K\"ahler potential, whilst the gauge fields it sources---characterized in turn by the magnetic and electric charges $(p^{A}{}', q_A')$---fix the flux quanta to be
\begin{equation}\label{eq:BHcharges}
p^A{}' = \frac{1}{4\pi} \int_{\mathbf{S}^2} F^A\, , \qquad q_A' = \frac{1}{4\pi} \int_{\mathbf{S}^2} G_A \, , \qquad G_A^- = \bar{\mathcal{N}}_{AB} F^{B,-}\, .
\end{equation}
Eq.~\eqref{eq:BHcharges} admits the following solution for the $U(1)$ field strengths \cite{Castellano:2025yur}
\begin{equation}\label{eq:backgroundfieldsBH}
 R^{2}\, F^A = p^A{}' \, \omega_{\mathbf{S}^2} - 2 \text{Re}\, (C X^A) \, \omega_{\mathrm{AdS}_2} \, , \qquad
 R^{2}\, G_A = q_A' \, \omega_{\mathbf{S}^2} - 2\text{Re}\, (C \mathcal{F}_A) \, \omega_{\mathrm{AdS}_2} \, ,
\end{equation}
where
\begin{equation}\label{eq:attvalues}
C X^A = \text{Re}\, (C X^A) + \frac{i}{2} p^A{}' \, , \qquad C \mathcal{F}_A = \text{Re}\, (C \mathcal{F}_A) + \frac{i}{2} q_A' \, ,
\end{equation}
correspond to the stabilized complex moduli determined via the attractor mechanism \cite{Ferrara:1995ih,Strominger:1996kf,Ferrara:1996dd,Ferrara:1996um}. These consist of a set of algebraic equations of the form (for more details on our conventions see \cite{Castellano:2025ljk})
\begin{equation}
\, p^A{}' = 2 \,\mathrm{Im}\,(C X^A)\,, \qquad q_A' = 2\, \mathrm{Im}\,(C \mathcal{F}_A)\,.
\end{equation}

Additionally, in four-dimensional $\mathcal{N}=2$ theories, the mass of a BPS point-like object (expressed in Planck units) is given by
\begin{equation}\label{eq:BPSmass}
m= |Z|\, ,\qquad Z= e^{K/2} \left(p^A \mathcal{F}_A-q_A X^A\right)\, ,
\end{equation}
with $(p^A, q_A)$ denoting the charges of the particle. Its dynamics within global AdS$_2\times \mathbf{S}^2$, parametrized by $(\tau, \chi, \theta,\phi)$ as in eq.~\eqref{eq:globalmetricAdS2}, can be obtained from the corresponding 1d action. The latter takes, in the bosonic sector, the following form \cite{Claus:1998ts,Billo:1999ip, Castellano:2025yur}\footnote{To write the gauge interaction as in the right hand side of eq.~\eqref{eq:globalwordlineactionAdS2xS2} one needs to substitute in \eqref{eq:backgroundfieldsBH} the volume 2-forms corresponding to the $\mathrm{AdS}_2$ and $\mathbf{S}^2$ factors, which are given by
\begin{equation}
 \omega_{\rm{AdS}_2} = {R^2}\cosh\chi\, d{\tau}\wedge d\chi \, , \qquad  \omega_{\mathbf{S}^2} = R^2 \sin \theta \, d\theta \wedge d\phi\,.
\end{equation}}
\begin{equation}\label{eq:globalwordlineactionAdS2xS2}
 S_{wl} = - \int_\gamma d\sigma \left[2mR \sqrt{\cosh^{2}{\chi} \ \dot{\tau}^2-\dot{\chi}^2 -\dot{\theta}^2-\sin^2\theta \,\dot{\phi}^2} + \left( -q_e\, {\dot{\tau}} \, \sinh\chi \,+ q_m \cos \theta\, \dot{\phi}\right)\right]\, ,
\end{equation}
with
\begin{equation}\label{eq:qeqm}
q_e = 2 \, \mathrm{Re}\,(\bar{Z}_{\rm BH} Z) \, , \qquad q_m = 2 \,  \mathrm{Im}\,(\bar{Z}_{\rm BH} Z) = p^A \, q_A' -  q_A \, p^A{}'\, .
\end{equation}
Here, $\gamma$ denotes the trajectory of the particle in spacetime, and we introduced the notation $\dot{x}^\mu := \frac{dx^\mu}{d\sigma}$, where $\sigma$ represents an arbitrary parameter along the worldline.

\medskip

An important comment worth making at this point concerns the coefficient appearing in front of the kinetic term in the worldline action, which may be expressed as
\begin{equation}
\tilde{m} := 2 |Z| R = 2 |\bar{Z}_{\rm BH} Z|\, ,
\end{equation}
where we substituted the explicit value of the AdS$_2$ radius given in eq.~\eqref{eq:BHcentralcharge}. Consequently, for any BPS particle propagating near the black hole horizon the following relation holds \cite{Castellano:2025yur}
\begin{subequations}
\begin{equation}\label{eq:quadconstraint}
q_e^2 + q_m^2 = \tilde{m}^2\, ,
\end{equation}
\begin{equation}\label{eq:upperboundchargetomass}
|q_e| = 2 \left| \mathrm{Re}\,(\bar{Z}_{\rm BH} Z) \right| \leq 2  \, |\bar{Z}_{\rm BH} Z| = \tilde{m}\, ,
\end{equation}
\end{subequations}
with equality achieved in \eqref{eq:upperboundchargetomass} exactly for the extremal limit, i.e., when $q_m = 0$. 

\subsubsection{Equations of motion and spacetime trajectories}\label{sss:EOMs}

To analyze the classical paths that extremize the action functional displayed in eq.~\eqref{eq:globalwordlineactionAdS2xS2}, we introduce an einbein field $h(\sigma)$ which allows us to rewrite it as
\begin{equation}
 S_{wl} = \frac12 \int_\gamma d\sigma\left[ h^{-1}\left(-\cosh^{2}{\chi} \ \dot{\tau}^2 + \dot{\chi}^2 +\dot{\theta}^2 + \sin^2\theta \,\dot{\phi}^2\right)-h\tilde{m}^2 + q_e\, {\dot{\tau}} \, \sinh\chi \,- q_m \cos \theta\, \dot{\phi}\right]\, .
\end{equation}
Using the worldline reparametrization symmetry, one can choose locally $h(\sigma)=1$, provided we simultaneously impose the on-shell constraint
\begin{equation}\label{eq:apHamiltonianconst}
  H= \frac{1}{2} 
\left[ p_\chi^2 - \left( p_\tau \, \text{sech}{\chi} - q_e \, \tanh{\chi} \right)^2\right] + \frac12 p_\theta^2 + \frac12\csc^2\theta\, \left( p_\phi+q_m \cos \theta\right)^2 \stackrel{!}{=} -\frac{\tilde{m}^2}{2}\, ,
\end{equation}
where we defined above the momenta canonically conjugate to the embedding coordinates
\begin{equation}\label{eq:conservedchargesglobalAdS}
p_\chi = \dot{\chi}\, ,\qquad
p_{\tau} = -\dot{\tau}\, \cosh ^2{\chi}+{q_e} \sinh \chi\, ,\qquad
p_\theta =  \dot{\theta}\, ,\qquad
p_\phi = \sin^2\theta\, \dot{\phi}-q_m \cos \theta\, .
\end{equation}
The conserved Noether charges associated to invariance under $\tau$ and $\phi$ shifts are the angular momentum ($j$) and energy ($E$) per unit mass
\begin{equation}
j =p_\phi\, ,\qquad
E= -p_\tau\, ,
\end{equation}
whilst the equations of motion for the remaining $(\chi, \theta)$-coordinates read
\begin{subequations}\label{eq:thetapsieoms}
\begin{align}
\dot{p}_\chi &=  \ddot\chi = q_e \dot{\tau} \, \cosh{\chi} - \dot{\tau}^2 \, \sinh{\chi} \, \cosh{\chi}  = \text{sech}^3{\chi} \,(p_\tau -{q_e} \sinh{\chi}) (p_\tau \sinh{\chi} + q_e)   \, , \label{eq:eompsi}\\
\dot{p}_\theta &= \ddot \theta=  \sin\theta\left( \cos \theta\,\dot{\phi}^2+q_m  \dot \phi\right)= \dot{\phi} \tan \theta \left(\dot{\phi}-j\right)\, .
\end{align}
\end{subequations}
These must be supplemented with the Hamiltonian constraint \eqref{eq:apHamiltonianconst} which, after solving for the sphere dynamics \cite{Castellano:2025yur}, can be conveniently expressed as
\begin{equation}\label{eq:effectiveradialpoteglobal}
 p_\chi^2 + V(\chi) =0\, ,\qquad V(\chi)=  m_{\rm eff}^2-(E\, \text{sech}{\chi} + {q_e} \tanh{\chi})^2\,,
\end{equation}
with $m_{\rm eff}^2=\tilde{m}^2+\ell^2$ being the effective 2d mass, thereby incorporating the inertia associated to the orbital angular momentum $\ell =\sqrt{j^2-q_m^2}$ along $\mathbf{S}^2$. We have shown the behavior of the radial potential---depending on the electric charge-to-mass ratio $q_e/m_{\rm eff}$---in Figure \ref{fig:RadialPotglobal} below. 

\begin{figure}[t!]
\centering
\subcaptionbox{\label{sfig:subextremalglobal}}{\includegraphics[width=0.45\linewidth, height=4cm]{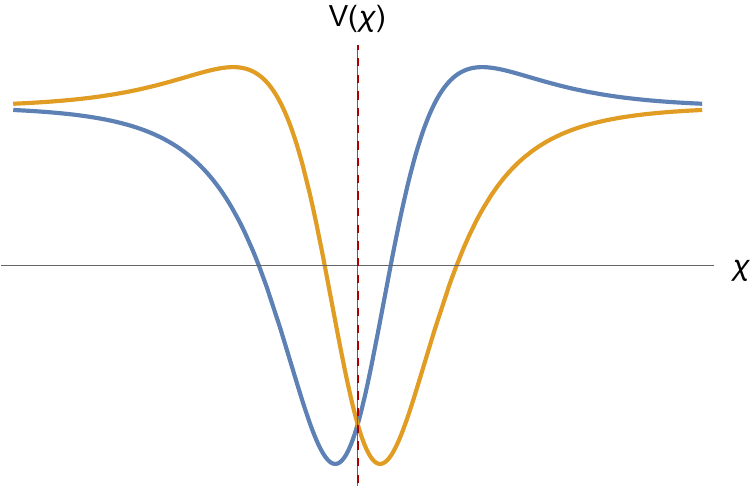}}
\hspace{0.5cm}
\subcaptionbox{\label{sfig:supextremalglobal}}{\includegraphics[width=0.45\linewidth, height=4cm]{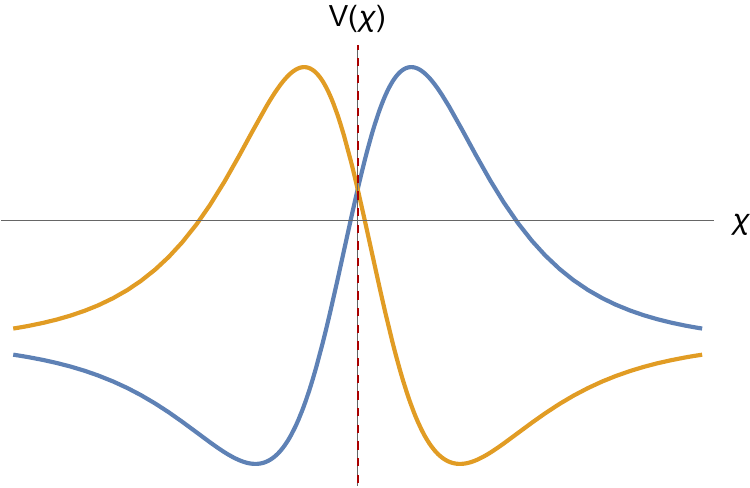}}
\vspace{1cm}
\subcaptionbox{\label{sfig:extremalglobal}}{\includegraphics[width=0.45\linewidth, height=4cm]{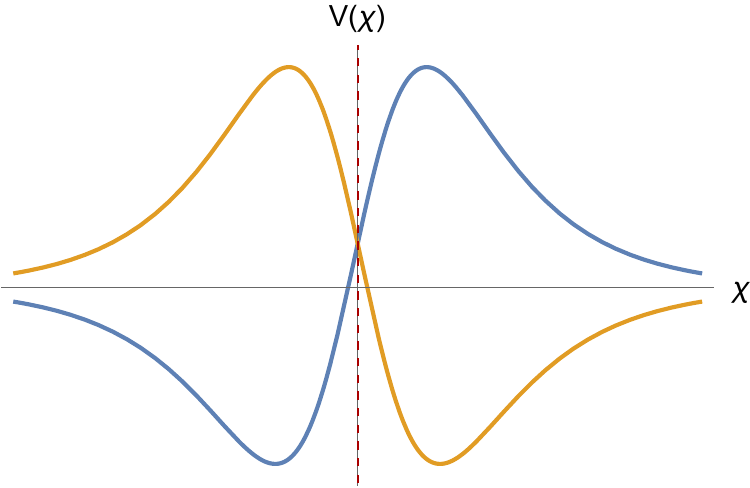}}
\caption{\small Effective potential $V(\chi)$ controlling the radial dynamics in global AdS$_2 \times \mathbf{S}^2$, cf. eq.~\eqref{eq:effectiveradialpoteglobal}. The dashed vertical line denotes the `center' of Anti-de Sitter space at $\chi=0$. The qualitative features of the potential depend on whether $\textbf{(a)}$  $q_e^2 <\tilde{m}^2 + \ell^2$ (subextremal), $\textbf{(b)}$ $q_e^2 >\tilde{m}^2+ \ell^2$ (superextremal), or $\textbf{(c)}$ $q_e^2 =\tilde{m}^2 +\ell^2$ (extremal). We show the corresponding effective potential for both the particle ($E q_e>0$, yellow) and its CPT conjugate ($E q_e<0$, blue).}
\label{fig:RadialPotglobal}
\end{figure}

\medskip

\begin{figure}[t!]
    \centering
    {\subcaptionbox{\label{a}}{
	\includegraphics[width=0.45\linewidth, height = 7.5cm]{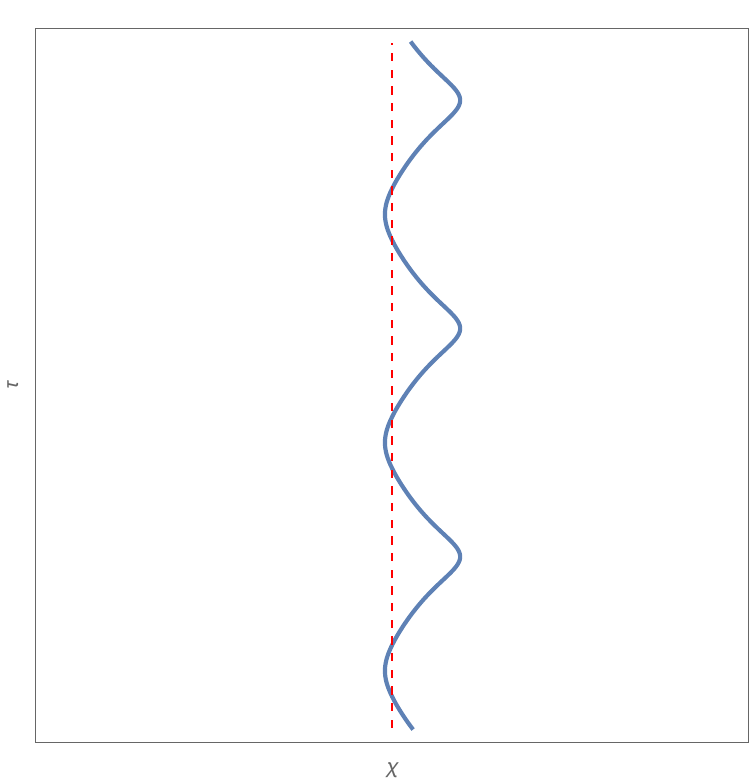}}
        \hspace{0.6cm}
    \hspace{0.1cm}}
    {\subcaptionbox{\label{b}}{
	\includegraphics[scale=0.4]{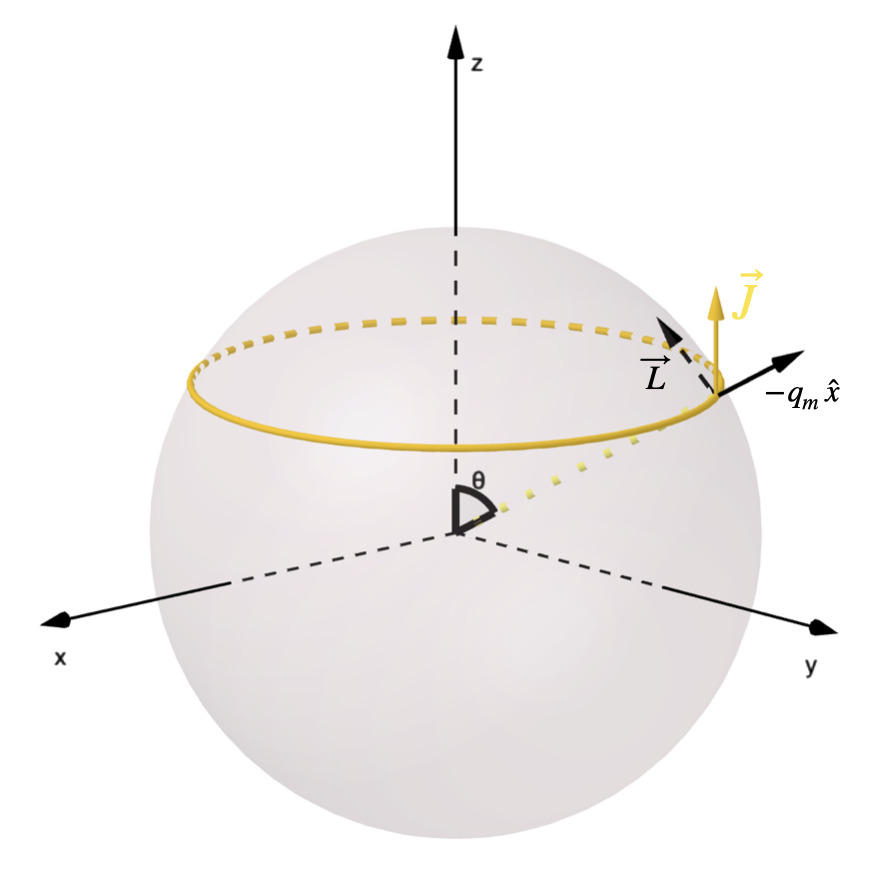}}}
		\caption{\small Depiction of the spacetime trajectories associated to charged subextremal particles in 4d $\mathcal{N}= 2$ AdS$_2 \times \mathbf{S}^2$ geometries. The dynamics along the sphere (right) is controlled by the generalized angular momentum vector $\boldsymbol{J}$, around which the particle precesses, whereas in AdS$_2$ (left) particles are confined within some finite distance from the conformal boundaries and exhibit periodic motion.} 
		\label{fig:intrinsicmotion}
\end{figure}

However, instead of directly integrating the equations of motion \eqref{eq:thetapsieoms}, one may obtain the form of the intrinsic trajectories followed by charged BPS particles in AdS$_2 \times \mathbf{S}^2$ upon exploiting the symmetries exhibited by the system. Indeed, from an algebraic perspective, turning on some constant and everywhere orthogonal gauge fields along both Anti-de Sitter and the 2-sphere preserves the isometries of the underlying 4d spacetime \cite{Comtet:1986ki, Dunne:1991cs, Pioline:2005pf}. Those are identified with conformal and rotational transformations of the form $SU(1,1) \times SU(2)$, which can be encoded, in turn, into the superconformal group $SU(1,1|2)$. In particular, the generators for the aforementioned (bosonic) subalgebras read---in global coordinates---as
\begin{equation}\label{eq:SU11generators}
\begin{aligned}
  K_0 &= -\text{sech}\chi \, \sin\tau \, (p_\tau \sinh\chi +q_e)+p_\chi \cos\tau \,,\\
  K_\pm &= \mp p_\chi \sin \tau \mp \text{sech}\chi \cos \tau (p_\tau \sinh \chi+q_e)+p_\tau\,,
\end{aligned}
\end{equation}
for $\mathfrak{su}(1,1)$, whereas in the case of $\mathfrak{su}(2)$ one has
\begin{equation}\label{eq:SU2generators}
\begin{aligned}
    J_\pm =  \pm i e^{\pm i\phi} \left[ p_\theta \pm i \left( \cot \theta\, p_\phi+q_m \csc \theta\right)\right]\, ,\qquad J_0 = p_\phi\, .
    \end{aligned}
\end{equation}
These quantities are readily seen to satisfy the commutator relations
\begin{equation}\label{eq:SU2xSL2algebra}
\begin{aligned}
    &\big \{ J_+, J_- \big\}_{\rm PB} = 2J_0\, , \quad \big \{ J_0, J_\pm \big\}_{\rm PB} = \pm J_\pm\, ,\\
    &\big \{ K_+, K_- \big\}_{\rm PB} = -2K_0\, , \quad \big \{ K_0, K_\pm \big\}_{\rm PB} = \pm K_\pm\, ,\\
    &\big \{ J_i, K_j \big\}_{\rm PB}=0\, ,
\end{aligned}
\end{equation}
with respect to the familiar Poisson bracket, given by
\begin{equation}\label{eq:Poissonbracketdef}
\begin{aligned}
    \big \{ A(q,p), B(q,p) \big\}_{\rm PB}= \frac{\partial A}{\partial q^i}\frac{\partial B}{\partial p_i}-\frac{\partial B}{\partial q^i}\frac{\partial A}{\partial p_i}\, ,
    \end{aligned}
\end{equation}
where $A(q,p), B(q,p)$ denote an arbitrary pair of functions defined on phase space. However, one cannot freely choose the conserved charges, as they are not fully independent. Indeed, the mass-shell constraint \eqref{eq:apHamiltonianconst} admits the following group-theoretic expression \cite{Castellano:2025yur}
\begin{equation}\label{eq:Casimirconstraint}
   C_2^{\mathbf{S}^2} + C_2^{\text{AdS}_2}=0\, ,
\end{equation}
thereby linking the quadratic Casimirs of both $SU(1,1)$ and $SU(2)$, which depend on the generalized charges according to
\begin{equation}
   C_2^{\mathbf{S}^2}=J_0^2 +\frac12 \left( J_+J_- + J_-J_+ \right)\, ,\qquad C_2^{\text{AdS}_2}=K_0^2 -\frac12 \left( K_+K_- + K_-K_+ \right)\, .
\end{equation}
Consequently, the motion on the sphere may be easily deduced by asking for the generalized angular momentum vector $\boldsymbol{J}$ to be aligned with the $J_0$ direction---possibly after some $SU(2)$ rotation. This implies $J_+=J_-=0$, hence imposing the dynamical condition (cf. eq.~\eqref{eq:SU2generators})
\begin{equation}\label{eq:dynamicalcondS2}
    \qquad p_\phi=j,\,\qquad \cos \theta=-\frac{q_m}{j}\, .
\end{equation}
Similarly, the trajectories within AdS$_2$ can be obtained by solving the implicit equation
\begin{equation}\label{eq:trajectoryeqAdS2}
    (K_+ - K_-) \cos{\tau} + 2 K_0 \sin{\tau} = -2 q_e \, \text{sech}{\chi} - 2 p_\tau \tanh{\chi}\, ,
\end{equation}
that follows directly from the definition of the $SU(1,1)$ charges. From this, one concludes that the dynamics along Anti-de Sitter becomes periodic in time ($\tau \sim \tau +2\pi$), see Figure \ref{fig:intrinsicmotion}.

\subsubsection{Static and stationary paths}\label{sss:static&stationarypaths}

To close this section, we want to investigate certain special trajectories that are singled out by the symmetries of the underlying theory. More concretely, given that both the background \cite{Gibbons:1984kp,Kallosh:1992gu,Kallosh:1992ta,Ferrara:1997yr} and the particles \cite{Ceresole:1995ca} under consideration preserve some amount of supersymmetry, it is natural to ask whether some of the previously described geodesics might themselves be supersymmetric.

\medskip

In what follows, we describe a class of stationary configurations in AdS$_2 \times \mathbf{S}^2$, where the particle rests at an equilibrium position---determined by its charges---in Anti-de Sitter while simultaneously precessing around the sphere due to its angular momentum. Later, in Sections \ref{s:staticprobes} and \ref{s:stationaryprobes}, we show that these solutions preserve exactly half of the background supersymmetries, making them $\frac12$-BPS. This encompasses the fully static paths analyzed in \cite{Simons:2004nm}, where supersymmetry was verified via a worldline $\kappa$-symmetry analysis, and further generalizes them to cases with non-zero angular momentum on the sphere.

To illustrate the simplest instance, we begin with the static case. In terms of the global coordinates introduced in \eqref{eq:globalmetricAdS2}, these trajectories correspond to constant values of $(\chi, \theta, \phi)$, implying that the motion is characterized by having $\ell = 0$. Notice that the precise location along the sphere determines the direction of the generalized angular momentum vector, which only receives contributions from the electromagnetic field
\begin{equation}
    \boldsymbol{J}=-q_m\, (\sin\theta \cos \phi,\,\sin \theta \sin \phi,\, \cos \theta)\, .
\end{equation}
On the other hand, for the particle to remain at fixed $\chi$, the minimum exhibited by the effective potential \eqref{eq:effectiveradialpoteglobal} needs to be such that $V(\chi_{\rm min})=0$ (cf. Figure \ref{sfig:subextremalglobal}). The latter occurs for $\sinh\chi=q_e/E$, and thus having $p_\chi=0$ at all times requires the energy to be
\begin{equation}\label{eq:staticenergy}
    E=\sqrt{\tilde{m}^2-q_e^2} = |q_m|\, ,
\end{equation}
where in the last step we made use of \eqref{eq:quadconstraint}.

Alternatively, from eq.~\eqref{eq:trajectoryeqAdS2} we may directly deduce that in order to have a genuine static trajectory in AdS$_2$ one has to choose the $SU(1,1)$ conserved charges as follows
\begin{equation}
    K_0 = 0\,,\qquad K_+ = K_-\,.
\end{equation}
This, when combined with the Hamiltonian constraint \eqref{eq:apHamiltonianconst}, implies a precise relation between the charges and the energy of the probe
\begin{equation}
    K_+^2 =q_m^2\,,
\end{equation}
thereby forcing the equilibrium position to happen at
\begin{equation}\label{eq:dynamicalcondstatic}
    \sinh{\chi} = \frac{\cos (CZ)}{|\sin (CZ)|}\, ,
\end{equation}
in agreement with our previous considerations. 
\begin{figure}[t!]
   \centering
{\includegraphics[width=0.5\linewidth, height=5cm]{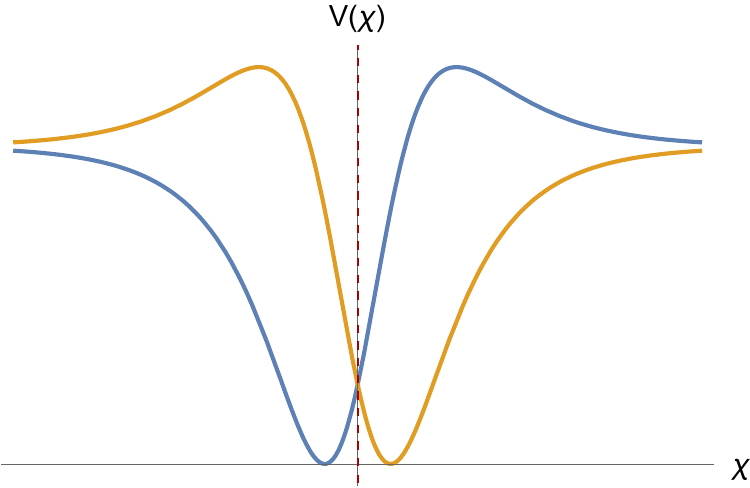}} 
\caption{\small Whenever the global energy of the BPS probe reaches certain minimum value, i.e., for $E=\sqrt{j^2}$, the on-shell trajectory becomes stationary in AdS$_2\times\mathbf{S}^2$, such that the particle stays at a constant radial distance from the boundary determined by its conserved charges. The resulting effective potential therefore exhibits a minimum at $\chi_{\rm min}=\text{sinh}^{-1} (q_e/|j|)$ that verifies $V(\chi_{\min})=0$. The yellow (blue) line indicates a charged particle with $p_\tau \, q_e <0$ ($p_\tau \, q_e >0$). }
\label{fig:minPotentialSUSY}
\end{figure}

\medskip

On a similar note, it is also possible to obtain stationary trajectories with non-zero angular momentum $\ell$ along the internal $\mathbf{S}^2$. The analysis proceeds analogously to the static case above, with only minor modifications (see \cite{Castellano:2025yur} for additional discussion).
In particular, the radius of the configuration is now modified to
\begin{equation}\label{eq:radialequilibriumangmom}
   \csch \chi = \frac{\sqrt{j^2}}{q_e}\, , \qquad j^2 = q_m^2 + \ell^2 \,,
\end{equation}
indicating that, as $\ell$ increases, the equilibrium position moves toward the center of AdS$_2$. As a result, just as in the static solution, the effective potential in the AdS radial direction still develops a global minimum at \eqref{eq:radialequilibriumangmom} satisfying $V(\chi_{\rm min})=0$ when $E=\sqrt{j^2}$, see Figure \ref{fig:minPotentialSUSY}. 

\medskip

\noindent In the remainder of this note, we dedicate our efforts to proving that the special trajectories described herein are indeed supersymmetric. To that end, we present a  $\kappa$-symmetry argument that recovers the dynamical conditions \eqref{eq:dynamicalcondS2} and \eqref{eq:radialequilibriumangmom} by requiring some supercharges to remain unbroken. We also show that these configurations saturate a BPS-like bound.\footnote{This can be anticipated as well from the Casimir constraint $K_+K_--K_0^2=j^2$. Indeed, upon substituting the identity $K_+K_-= \frac14 \left[(K_++K_-)^2-(K_+-K_-)^2\right]$ and using $E^2=\frac14 (K_++K_-)^2$, one finds $E^2 \geq j^2$.}

\section{Supersymmetric Static Probes}\label{s:staticprobes}

In this section, we review in detail the proof of \cite{Simons:2004nm} that certain static D-brane probe configurations preserve half of the enhanced supercharges in the near-horizon black hole background. Hence, after briefly introducing both supersymmetry and $\kappa$-symmetry on the worldline theory (cf. Section \ref{ss:susy&kappasym}), we proceed in Section \ref{ss:Killingspinors} to discuss the explicit Killing spinors associated to the spacetime solution, as obtained in \cite{Alonso-Alberca:2002wsh}. Finally, Section \ref{ss:recoveringStrominger} is devoted to providing the details of the algebraic argument, drawing attention to various salient features that will be important for later generalizations.

\subsection{Supersymmetric trajectories}\label{ss:susy&kappasym}

\subsubsection{$\kappa$-symmetry and worldvolume supersymmetry restoration}

The procedure for determining whether a supergravity solution preserves supersymmetry is by now well-established. In practice, one must identify the unbroken supercharges (if any) that generate transformations leaving the full configuration invariant. More concretely, one first considers the corresponding infinitesimal variations involving all dynamical fields altogether, then imposes that they must vanish, and finally determines when the resulting conditions can be solved simultaneously. Furthermore, since the supersymmetry operation acting on bosonic (fermionic) fields is linear in the fermions (bosons), purely bosonic backgrounds yield trivial conditions from the bosonic sector. As a result, it suffices to focus on the variations associated to the fermions, which give rise to the so-called Killing spinor equations (KSEs). The solutions to these equations are referred to as Killing spinors.

\medskip

Here, we are interested in studying the motion of BPS particles in purely bosonic backgrounds and, in particular, we want to determine which are their possible supersymmetric trajectories. To do so, we will adopt the probe approximation point of view, i.e., we consider the worldvolume action describing a wrapped D-brane propagating in a \emph{fixed} background geometry (see \cite{Simon:2011rw} for a comprehensive review). There, spacetime supersymmetry can be encoded using the superspace formalism in terms of the transformation law of the target space coordinates $X^\mu$ and their fermionic Grassmann-valued partners $\Theta$. The latter are simply superspace translations of the form $\delta_\epsilon \Theta = \epsilon$, with $\epsilon$ denoting some spacetime spinor of the same kind. However, notice that by fixing the spacetime trajectory followed by the brane probe we necessarily break all the supersymmetries, since they always induce a non-vanishing transformation on the fermions. 

At the same time, the worldline theory thus obtained usually admits an additional gauge freedom called $\kappa$-symmetry \cite{Becker:1995kb, Bergshoeff:1997kr}, which acts on the fermions as a half-rank projection $\delta_\kappa \Theta = (\mathbf{1} + \Gamma) \kappa$. Here, $\kappa$ is a local Grassmann parameter whereas $\Gamma$ defines a traceless involution that depends on the details of the theory under consideration. Supersymmetry is then restored if the aforementioned global transformation parametrized by $\epsilon$ can be compensated via some $\kappa$-variation. Consequently, to determine the amount of unbroken supercharges left by the particle one needs to solve the algebraic condition
\begin{equation} \label{eq:kappaSymCondition1}
    0 = \delta_\epsilon \Theta + \delta_\kappa \Theta \,,
\end{equation}
where $\epsilon$ is a Killing spinor of the bosonic background. Exploiting the orthogonality of the projectors $P_\pm=\frac12(\mathbf{1}\pm \Gamma)$, eq.~\eqref{eq:kappaSymCondition1} can be conveniently written as
\begin{equation} \label{eq:kappaSymCondition2}
    (\mathbf{1} - \Gamma) \, \epsilon = 0 \,.
\end{equation}

\subsubsection{BPS particles in 4d $\mathcal{N}=2$ backgrounds}

The $\kappa$-symmetry projector \eqref{eq:kappaSymCondition2} for a BPS particle moving in a background solution of 4d $\mathcal{N}=2$ supergravity coupled to $n_V$ vector multiplets takes the explicit form
\begin{subequations} \label{eq:conditionKappaSusy}
\begin{align}
    & \epsilon_A + i \, e^{i\alpha}\, \Gamma_\kappa \, \epsilon_{AB} \, \epsilon^B  = 0 \,, \\[2mm]
    & \epsilon^A + i \, e^{-i\alpha} \, \Gamma_\kappa \, \epsilon^{AB} \, \epsilon_B  = 0 \, .
\end{align}
\end{subequations}
Above, $\epsilon_A = (\epsilon^A)^*$ represents a Killing spinor expressed in the Weyl representation, with $A = 1,2$, labeling the two underlying supersymmetries; $\epsilon_{AB}$ corresponds to the Levi--Civita symbol, $\alpha$ is the complex phase of the central charge associated to the BPS particle, namely
\begin{equation}\label{eq:phasecentralcharge}
    Z = |Z| \, e^{i \alpha}\,,
\end{equation}
and $\Gamma_\kappa$ gives the projection of the gamma matrices $\gamma^a$ onto the particle worldine $X^\mu (\tau)$
\begin{equation}\label{eq:kappasymmprojector}
    \Gamma_\kappa = \gamma_a \, e_\mu {}^a \, \dot{X}^\mu \frac{1}{\sqrt{-h_{\tau \tau}}} \,, \qquad h_{\tau \tau} =  \dot{X}^\mu \dot{X}^\nu g_{\mu \nu} \,.
\end{equation}
Here, $g_{\mu \nu}$ is the spacetime metric and $ e_\mu {}^a$ denote the associated vierbeins. Using instead Majorana spinors (cf. eq.~\eqref{eq:AppMapToMajorana}) and expressing  $\epsilon_{IJ} = i (\sigma^2)_{ij}$, we can rewrite \eqref{eq:conditionKappaSusy} as
\begin{equation} \label{eq:kappaMaioranaImaginaryOriginal}
    \epsilon =  e^{- i \alpha \gamma_5} \, \Gamma_\kappa \, \sigma^2  \, \epsilon \,,
\end{equation}
where $\epsilon$ is a doublet of Majorana fermions.

\noindent The previous relation was originally determined in \cite{Billo:1999ip} upon considering the supersymmetric extension of the worldline action \eqref{eq:globalwordlineactionAdS2xS2}, which is given by
\begin{equation}
 S_{wl} = -2  \int_{\gamma} |Z| \left[(- \Pi^a V^b + \frac{1}{2} \Pi^a \Pi^b e) \eta_{ab} + \frac{1}{2}e\right] + \int_\Sigma (p^A G_A - q_A F^A) \,, 
\end{equation}
with $\Sigma$ any 2d surface such that $\gamma=\partial \Sigma$, and requiring that the (off-shell) supersymmetry transformations of the spacetime bosons leave the action invariant \cite{Andrianopoli:1996cm}. Here, $V^a$ is the spacetime supervierbein, $e$ refers to the worldline einbein 1-form, and $\Pi^a$ denote some auxiliary 0-forms that on-shell become the pull-back of the supervierbein onto the worldline $\gamma$.

\subsection{Killing spinors}\label{ss:Killingspinors}

In this section, we focus on solving the KSEs in the same setup considered in Section \ref{s:review}, i.e., with the metric and gauge backgrounds arising as near-horizon limits of BPS black hole solutions in 4d $\mathcal{N}=2$ supergravity. We start by showing that for this class of geometries the only non-trivial KSE is the one associated with the gravitino. Subsequently, we review the method of \cite{Alonso-Alberca:2002wsh} to construct Killing spinors in symmetric coset spaces and we apply it to the case of AdS$_2 \times \mathbf{S}^2$. The final result we get is the following expression
\begin{equation} 
\epsilon (x) =  e^{-\frac{1}{2}  \chi \gamma^0 \sigma^2} e^{\frac{1}{2} \tau \gamma^1 \sigma^2} e^{- \frac{1}{2}(\theta - \pi/2)  \gamma^0 \gamma^1 \gamma^2 \sigma^2} e^{-  \frac{1}{2 } \phi \gamma^0 \gamma^1 \gamma^3 \sigma^2 } \epsilon_0\, . 
\end{equation}

\subsubsection{KSEs for AdS$_2 \times\mathbf{S}^2$}

The superfield content of 4d $\mathcal{N}=2$ effective field theories (EFTs) include the supergravity multiplet $\{e^a{}_\mu, \psi^I{}_\mu, A_\mu\}$, $n_{V}$ vector multiplets $\{A^i{}_\mu, \lambda^{iA},z^i\}$, and $n_{H}$ hypermultiplets $\{q^u, \zeta_\alpha\}$. 
The Killing spinors $\epsilon_A$ of a purely bosonic background are the solutions of the KSEs \cite{Andrianopoli:1996cm,Ferrara:1997yr, Ortin:2015hya} 
\begin{subequations} \label{eq:KSEs}
\begin{align}
    \delta_\epsilon \psi_{\mu A} & = \nabla_\mu \epsilon_A + {\frac12}\epsilon_{AB} W^{-}_{\mu\nu} \gamma^\nu \epsilon^B = 0 \,, \\[2mm]
     \delta_\epsilon \lambda^{i A } & = i \gamma^\mu\partial_\mu z^i \epsilon^A  +\frac{i}{4} \mathcal{F}_{\mu \nu}^{-\,i}\gamma^{\mu \nu} \epsilon_B \epsilon^{AB} = 0 \,, \\[2mm]
     \delta_\epsilon \zeta_\alpha & = i C_{\alpha\beta}\, \mathcal{U}_u^{B \beta} \gamma^\mu\partial_\mu q^u \epsilon^A \epsilon_{AB} = 0 \,,
\end{align}
\end{subequations}
where $C_{\alpha \beta}$ is the $Sp(2n_H)$-invariant metric, $\mathcal{U}_u^{B \beta}$ are the so called quadbein---i.e., the vielbein of the quaternionic K\"ahler space,
$W_{\mu\nu}^{-}$ is  the graviphoton field strength and $\mathcal{F}^{-\,i}_{\mu\nu}$ refer to the linear combination of abelian fields belonging to the vector multiplets. The latter two can be related to the quantities introduced in Section \ref{ss:geodesics} via 
\begin{equation}\label{eq:GravPhotAndF}
\begin{aligned}
    W^- & =  e^{K/2} \left( \mathcal{F}_AF^{A,\,-}-X^AG_A^- \right) \,, \\[2mm]
    D_i \left(e^{K/2} X^A \right) \mathcal{F}^{-\,i}_{\mu\nu} & = - i \left(F^{A, -}_{\mu\nu} - i e^{K/2}\bar{X}^A W^{-}_{\mu\nu} \right) \,. 
\end{aligned}
\end{equation}
Let us specialize \eqref{eq:KSEs} to the near-horizon limit of a BPS black hole. First, we notice that since all the scalars fields $z^i$ and $q^i$ are fixed at the attractor point, their derivatives vanish. Next, by susbtituting eqs.~\eqref{eq:BHcentralcharge},
\eqref{eq:backgroundfieldsBH} and \eqref{eq:attvalues} into \eqref{eq:GravPhotAndF} and using the spacetime metric
given in \eqref{eq:globalmetricAdS2}, one can readily verify that
\begin{equation}
    W^- = - \frac{i}{R} e^{i\varphi}\left({-}\omega_{\rm{AdS}_2}+i\, \omega_{\mathbf{S}^2}\right) \,, \qquad \mathcal{F}^{i, -} = 0 \,,
\end{equation}
where $\varphi$ is the phase of the black hole central charge 
\begin{equation}
    Z_{\rm BH} = | Z_{\rm BH}| \, e^{i \varphi} \,.
\end{equation}
The only non-trivial KSE is the one associated with the variation of the gravitino. Thus, writing the graviphoton field as\footnote{Our convention for the Hodge dual applied to a $p$-form field $A= \frac{1}{p!} A_{\mu_1\ldots\mu_p} dx^{\mu_1}\wedge \cdots \wedge dx^{\mu_p}$ is as follows
\begin{equation}
   \star A=\frac{\sqrt{-g}}{(4-p)! p!}\, A_{\mu_1\ldots\mu_p} \epsilon^{\mu_1\ldots\mu_p}_{\qquad\ \ \mu_{p+1}\ldots\mu_{4-p}}\,dx^{\mu_{p+1}}\wedge \cdots \wedge dx^{\mu_{4-p}}\,,\notag
\end{equation}
with the choice of orientation given by eq.~\eqref{eq:orientation}. From here, one finds that $\star\,\omega_{\mathbf{S}^2} = \omega_{\rm{AdS}_2}$ and $ \star\,\omega_{\rm{AdS}_2} = -\omega_{\mathbf{S}^2}$.}
\begin{equation} \label{eq:defF}
    W^- = i \, e^{i\varphi}\left(1 + i \,\star \right) F \,, \qquad F = \frac{{\omega}_{\rm{AdS}_2}}{R} \,,
\end{equation}
and using the identities \eqref{eq:AppRel1} and \eqref{eq:AppRel2}, we can easily determine the form of the corresponding KSE for Majorana spinors
\begin{equation}
  \nabla_\mu \epsilon + \frac{1}{2}e^{-i\varphi \gamma_5} \left(- F_{\mu\nu} \gamma^\nu - i\frac{\sqrt{-g}}{2} F_{\rho \sigma} \, \epsilon^{\rho \sigma}{}_{\mu\nu} \gamma^\nu \gamma_5 \right) \sigma^2 \epsilon =  0  \,.
\end{equation}
The above expression can be further simplified by means of an R-symmetry transformation. Indeed, the action of the global $U(1)$ subgroup on the graviphoton and the gravitino reads
\begin{equation}
    W^- \rightarrow e^{-i\beta} W^-  \,, \qquad \psi_{\mu} \rightarrow e^{\frac{i}{2}\beta \gamma_5} \psi_{\mu} \,,
\end{equation}
where the phase $\beta$ can be encoded into a complex rescaling of the fields $X^I$ and $\mathcal{F}_I$, which themselves induce a rotation on the Killing spinors $\epsilon$ 
\begin{equation}
    X^I \rightarrow e^{-i\beta} X^I \,, \qquad \mathcal{F}_I \rightarrow e^{-i \beta} \mathcal{F}_I \,, \qquad \epsilon \rightarrow e^{\frac{i}{2} \beta \gamma_5 } \epsilon \,.
\end{equation}
Hence, one can use the transformation above to set $\varphi = 0$ or, equivalently, to select the frame in which the black hole central charge is purely real. Upon doing so, we finally get
\begin{equation} \label{eq:finalFormKSE}
  \nabla_\mu \epsilon + \frac{1}{2} \left(- F_{\mu\nu} \gamma^\nu - i\frac{\sqrt{-g}}{2} F_{\rho \sigma} \, \epsilon^{\rho \sigma}{}_{\mu\nu} \gamma^\nu \gamma_5 \right) \sigma^2 \epsilon = 0  \,.
\end{equation}

\subsubsection{Killing spinors in symmetric spaces}

We now review the method put forward in \cite{Alonso-Alberca:2002wsh} to build solutions of the gravitino KSE that applies to those cases in which the target spacetime can be described as a homogeneous space of the form $G/H$. Throughout this section, we assume that the only supersymmetric condition we have to solve is the gravitino KSE and we refer to its solutions as Killing spinors.

\medskip

A symmetric manifold is an homogeneous space $G/H$ such that the algebras $\mathfrak{h}$ of $H$, $\mathfrak{l}$ of $G/H$ and $\mathfrak{g}$ of $G$ satisfy
\begin{equation}
    \mathfrak{g} = \mathfrak{h} \oplus \mathfrak{l} \,, \qquad [\mathfrak{h},\mathfrak{h}] \subset \mathfrak{h} \,, \qquad [\mathfrak{l},\mathfrak{h}] \subset \mathfrak{l} \,, \qquad [\mathfrak{l},\mathfrak{l}] \subset \mathfrak{h} \,.
\end{equation}
One of the (many) interesting aspects of this class of spaces is that out of a coset representative of $G/H$ we may readily build several other objects such as a vielbein basis and the associated connection 1-form. Let $T_I$, $M_i$ and $P_a$ denote the generators of $\mathfrak{g}$, $\mathfrak{h}$ and $\mathfrak{l}$, respectively, with indices split as $I = (a,i)$. Then, given a coset representative of $G/H$
\begin{equation}
    u(x) = e^{x^1 P_1} \dots e^{x^n P_n} \,,
\end{equation}
one can obtain the $\mathfrak{g}$-valued Maurer-Cartan 1-form
\begin{equation}\label{eq:MaurerCartan}
    V = u^{-1} d u = e^a P_a + \theta^i M_i \,.
\end{equation}
For the components $e^a$, we used the symbol usually reserved for the vielbein because they indeed correspond to those. The coefficients $\theta^i$ define instead a connection 1-form
\begin{equation}\label{eq:connection1form}
    \omega^a{}_b = \theta^i f_{ib}{}^a  \,,
\end{equation}
where $f_{IJ}{}^K$ are the structure constants of $\mathfrak{g}$. It is in fact simple to verify that \eqref{eq:MaurerCartan} satisfies the identity $ dV + V \wedge V = 0$ which, upon projection onto $\mathfrak{l}$, yields
\begin{equation}
    d e^a + (\theta^i f_{i b}{}^a ) \wedge e^b = 0 \,.
\end{equation}
Notice that $f_{ib}{}^a$ can be interpreted as the adjoint representation of $M_i$, such that \eqref{eq:connection1form} may be equivalently written as 
\begin{equation}
     \omega^a{}_b = \theta^i \, \Gamma_{\text{adj}}(M_i)^a{}_b \,.
\end{equation}

\medskip

Let us focus now on the gravitino KSE. It has the schematic form
\begin{equation}
    (\nabla_\mu + \Omega_\mu) \, \epsilon = 0 \,,
\end{equation}
such that contracting with the $dx^\mu$, one gets 
\begin{equation}\label{eq:KSEdiffform}
     \left(d - \frac{1}{4}\omega_{ab}\gamma^{ab} + \Omega\right) \, \epsilon = 0 \,.
\end{equation}
The spin connection in the second term can also be expressed in terms of the $\theta^i$ components of the Maurer-Cartan 1-form. In particular, we have
\begin{equation} \label{eq:SpinRepM}
    - \frac{1}{4}\omega_{ab}\gamma^{ab}  = \theta^i \, \Gamma_s(M_i)  \,,
\end{equation}
with the spinorial representation of the $M_i$ generators given by
\begin{equation}
    \Gamma_s (M_i) = -\frac{1}{4} f_{ib}{}^c \eta_{ca} \gamma^{ab} \,.
\end{equation}
In \cite{Alonso-Alberca:2002wsh} it was noted that in those cases in which $\Omega$ exhibits a structure
\begin{equation}
    \Omega = e^a \Gamma_s(P_a) \,,
\end{equation}
for some spinorial representation $\Gamma_s$ of the generators of $P_a$, the Killing spinor equation admits the following compact expression
\begin{equation}
     \bigg(d +  \Gamma_s(u)^{-1}\,  d \,\Gamma_s(u) \bigg) \, \epsilon = 0 \,,
\end{equation}
with
\begin{equation}
    \Gamma_s(u) = e^{x^1 \Gamma_s (P_1)} \dots e^{x^n \Gamma_s (P_n)} \,.
\end{equation}
The solutions to the Killing spinor equations then have the simple form
\begin{equation}
    \epsilon = \Gamma_s(u)^{-1} \epsilon_0 \,, 
\end{equation}
where $\epsilon_0$ is an arbitrary constant Majorana spinor.

\subsubsection{Killing spinors in AdS$_2 \times $S$^2$}

With the previous ingredients, we are now ready to construct the Killing spinors for any supersymmetric AdS$_2\times\mathbf{S}^2$ background of 4d $\mathcal{N}=2$ supergravity. Let us determine first the appropriate coset representatives. AdS$_2$ is equivalent to the coset space $SO(1,2)/SO(2)$, whereas $\mathbf{S}^2$ is isomorphic to the quotient $SO(3)/SO(2)$. The Lie algebras of these two spaces commute, allowing us to factorize the representative. Denoting by $P_0$, $P_1$ and $M_1$ the generators of $SO(1,2)$, and by $P_2$, $P_3$ and $M_2$ those of $SO(3)$, and normalizing them such that they satisfy
\begin{subequations}\label{eq:algebras}
\begin{align}
    & [P_0,P_1] = \frac{1}{R^2} M_1 \,, \qquad [M_1, P_0] = P_1 \,, \qquad [M_1,P_1] = P_0 \,, \\[1mm]
    & [P_2, P_3] = \frac{1}{R^2} M_2 \,, \qquad [M_2,P_2] = P_3 \,, \qquad [M_2,P_3] = - P_2 \,,
\end{align}
\end{subequations}
one obtains as coset representatives of $SO(1,2)/SO(2)$ and $SO(3)/SO(2)$, respectively, the following group elements
\begin{equation}\label{eq:initialChoice}
    u = e^{R \, x^0 P_0} e^{R \, x^1 P_1} \,, \qquad \tilde{u} = e^{R \, x^3 P_3} e^{R \, x^2 P_2} \,.
\end{equation}
On the other hand, to determine the relation between the parameters $x^\mu$ and our choice of coordinates \eqref{eq:globalmetricAdS2}, we can build the Maurer cartan 1-forms, extract the corresponding vierbein and compute the associated spacetime metric. After some manipulations, we get
\begin{subequations}
\begin{align}
   &  u^{-1} du = R \, dx^1 P_1 + R \, dx^0 \cosh x^1 P_0 + d x^0 \sinh x^1 M_1 \,, \\[1mm]
   & \tilde{u}^{-1} d\tilde{u} = R \, dx^2 P_2 + R \, dx^3 \cos x^2 P_3 -d x^3 \sin x^2 M_2  \,,
\end{align}
\end{subequations}
where we used the identities (valid for the algebra \eqref{eq:algebras})
\begin{subequations}
\begin{align}
& e^{-R \, s  \, P_1} P_0 \, e^{R \, s \, P_1} = \cosh{s} \, P_0 + \sinh{s} \,\frac{M_1}{R} \,, \\[1mm] 
& e^{- R \, s \, P_2} P_3  \, e^{R \, s \, P_2} = \cos{s} \, P_3 - \sin{s} \,\frac{M_2}{R}\,.
\end{align}
\end{subequations}
Hence, according to our logic before, we find the coordinate map 
\begin{equation}  \label{eq:mapXcoords}
    x^0 = \tau \,, \qquad x^1 = \chi \,, \qquad x^2 = \theta - \pi/2 \,, \qquad x^3 = \phi \,,
\end{equation}
leading to the coset representatives
\begin{equation} \label{eq:finalChoice}
    u = e^{R \, \tau P_0} e^{R \, \chi P_1} \,, \qquad \tilde{u} = e^{R \, \phi P_3} e^{R \, (\theta - \pi/2) P_2} \,,
\end{equation}
and the associated vierbein
\begin{equation}\label{eq:vierbeins}
    e^0 = R \, \cosh \chi d\tau  \,, \qquad e^1 = R \, d\chi \,, \qquad e^2 = R \, d\theta \,, \qquad e^3 = R \, \sin \theta d\phi  \,.
\end{equation}
Finally, we should determine the spinorial representation of the generators $P_a$. From eq.~\eqref{eq:finalFormKSE} we can read the explicit form of the $\Omega$ term in \eqref{eq:KSEdiffform}
\begin{equation}
    \Omega = e^a \left[ \frac{1}{2} \left(- F_{ab} \gamma^b - \frac{i}{2} F_{cd} \, \epsilon^{cd}{}_{ab} \gamma^b \gamma_5 \right) \sigma^2 \right]  \,,
\end{equation}
implying that the spinorial representation $\Gamma_s(P_a)$ is given by
\begin{equation}
    \Gamma_s(P_a) =  \frac{1}{2} \left(- F_{ab} \gamma^b - \frac{i}{2} F_{cd} \, \epsilon^{cd}{}_{ab} \gamma^b \gamma_5 \right) \sigma^2  \,,
\end{equation}
where $F = R^{-1} e^0 \wedge e^1$ was introduced in \eqref{eq:defF}. Explicitly, we obtain\footnote{One can easily extract the spinorial representation of $M_i$ using eq.~\eqref{eq:SpinRepM} an verify that the map $\Gamma_s$ respects the algebra \eqref{eq:algebras}.}
\begin{equation} \label{eq:spinorialReps}
    \begin{split}
    & \Gamma_s(P_0) = -\frac{1}{2R} \gamma^1 \sigma^2 \,,  \hspace{2cm} \Gamma_s(P_1) = \frac{1}{2R} \gamma^0 \sigma^2 \,, \\[2mm]
    & \Gamma_s(P_2)  = \frac{1}{2R} \gamma^0 \gamma^1 \gamma^2 \sigma^2 \,, \hspace{1.5cm} \Gamma_s(P_3) = \frac{1}{2 R} \gamma^0 \gamma^1 \gamma^3 \sigma^2 \,.
    \end{split}
\end{equation}
The Killing spinors are then
\begin{equation} \label{eq:KillingSpinorfinal}
    \epsilon = \Gamma_s\left(u \tilde{u}\right)^{-1} \epsilon_0 =  e^{-\frac{1}{2}  \chi \gamma^0 \sigma^2} e^{\frac{1}{2} \tau \gamma^1 \sigma^2} e^{- \frac{1}{2}(\theta - \pi/2)  \gamma^0 \gamma^1 \gamma^2 \sigma^2} e^{-  \frac{1}{2 } \phi \gamma^0 \gamma^1 \gamma^3 \sigma^2 } \epsilon_0 \,,
\end{equation}
as previously announced.
\subsection{Recovering the Simons-Strominger-Thompson-Yin result}\label{ss:recoveringStrominger}

Finally, we want to see which are the conditions we have to impose on the Killing spinors \eqref{eq:KillingSpinorfinal} in order to satisfy eq.~\eqref{eq:kappaMaioranaImaginaryOriginal}. We consider trajectories such that $\dot{\tau} = 1$, $\dot{\theta} = \dot{\chi} = \dot{\phi} = 0$. Hence, writing $\epsilon = v \, \epsilon'_0$, with $\epsilon'_0$ a new constant spinor and spacetime-dependent part $v(x)$
\begin{subequations}
    \begin{align}
          \epsilon_0' & = e^{- \frac{1}{2}(\theta - \pi/2)  \gamma^0 \gamma^1 \gamma^2 \sigma^2} e^{-  \frac{1}{2 } \phi \gamma^0 \gamma^1 \gamma^3 \sigma^2 } \epsilon_0 \,, \\[1mm]
          v & = e^{-\frac{1}{2}  \chi \gamma^0 \sigma^2} e^{\frac{1}{2} \tau \gamma^1 \sigma^2} \,,
    \end{align}
\end{subequations}
the $\Gamma_\kappa\,$-operator reduces to
\begin{equation}
    \Gamma_\kappa = \frac{1}{\sqrt{-h_{00}}}\,  e_\tau{}^0 \gamma_0  = -\gamma^0 \,,
\end{equation}
with the projected metric $h_{00} = -R^2 \cosh^2 \chi$. Using this, condition \eqref{eq:kappaMaioranaImaginaryOriginal} takes the form
\begin{equation} \label{eq:conditionStromingerStep1}
    \epsilon_0' = -v^{-1} e^{-i \alpha \gamma_5} \, \gamma^0 \, v \, \sigma^2  \epsilon_0' \,,
\end{equation}
whose r.h.s. can be written after some algebra (see Appendix \ref{app:identitiesABCD} for some useful identities) as 
\begin{equation} \label{eq:conditionStromingerStep3}
    \begin{split}
    v^{-1} e^{-i \alpha \gamma_5} \, \gamma^0 \, v \, \sigma^2 = & \; \cos \tau \left[ \cos \alpha \gamma^0 \sigma^2 + i \sin\alpha \sinh \chi \gamma_5 \right] \\[2mm]
    & + \sin \tau \left[ \cos\alpha \gamma^0 \gamma^1 + i \sin \alpha \sinh \chi \gamma_5 \gamma^1 \sigma^2 \right] \\[2mm]
    & - i \sin\alpha \cosh \chi \gamma_5 \gamma^0 \sigma^2 \,,
    \end{split}
\end{equation}
where we isolated terms which depend differently on $\tau$. Equation \eqref{eq:conditionStromingerStep1} can be satisfied only if we remove the spacetime dependence from \eqref{eq:conditionStromingerStep3}. This can be achieved by imposing first a condition on the AdS$_2$-independent part of the Killing spinor
\begin{equation} \label{eq:cond004}
    i \gamma_5 \gamma^0 \sigma^2 \epsilon_0' = \pm \epsilon_0' \,.
\end{equation}
Notice that this projection is compatible with the reality of the Majorana fermions because it satisfies the condition \eqref{eq:compatibilityCondition}. Interestingly, when written in terms of the constant part $\epsilon_0$ of the Killing spinor \eqref{eq:KillingSpinorfinal}, one finds that the appropriate projection that needs to be imposed is $\Omega(\theta, \phi) \, \epsilon_0=\pm \epsilon_0$, with\footnote{The matrix $\Omega(\theta, \phi)$ is an involution and thus defines a bona-fide projection, see discussion around eq.~\eqref{eq:compatibilityCondition}. Moreover, under the map $(\theta,\phi) \to (\pi-\theta,\phi+\pi)$, the projector gets reversed, namely $P_\pm=\frac12(1\pm \Omega(\theta,\pi))\to P_\mp$.}
\begin{equation}\label{eq:projectorstatic}
    \Omega(\theta, \phi)=\sin \theta \left( i\cos \phi \gamma_5\gamma^0\sigma^2 + \sin \phi \gamma^2 \gamma^0\right) + \cos \theta \gamma^3 \gamma^0 \,.
\end{equation}
The latter, of course, depends on the point of the sphere where the particle is placed, and if this corresponds to either one of the poles in $\mathbf{S}^2$, the operator reduces to $\Omega(\theta=0,\pi; \, \phi)={\pm} \gamma^3\gamma^0$. 

Therefore, substituting \eqref{eq:cond004} into eq.~\eqref{eq:conditionStromingerStep3}, we obtain
\begin{equation}
    \begin{split}
     v^{-1} e^{-i \alpha \gamma_5} \, \gamma^0 \, v \, \sigma^2 \epsilon_0'  = & \; \cos \tau \left[ \cos \alpha \mp \sin\alpha \sinh \chi \right] \gamma^0 \sigma^2\epsilon_0'   \\[2mm]
    & +\sin \tau \left[\cos \alpha \mp \sin\alpha \sinh \chi  \right]\gamma^0 \gamma^1 \epsilon_0'  \\[2mm]
    & \mp \sin\alpha \cosh \chi \epsilon_0'\, ,
    \end{split}
\end{equation}
which implies that it is possible to cancel the $\tau$-dependence by requiring
\begin{equation} \label{eq:cond001}
      \sin\alpha \sinh \chi = \pm \cos \alpha\,,
\end{equation}
or, equivalently,
\begin{equation}\label{eq:cond002}
    \cosh \chi = \frac{1}{|\sin \alpha|} \,.
\end{equation}
Thus, \eqref{eq:conditionStromingerStep1} is satisfied provided that 
\begin{equation} \label{eq:cond003}
   \pm \sin\alpha \cosh \chi = 1 \,.
\end{equation}
The compatibility between \eqref{eq:cond003} and \eqref{eq:cond002} fixes the specific projection condition we have to pick in \eqref{eq:cond004}. One finds 
\begin{equation}\label{eq:signprojectorstatic}
    i \gamma_5 \gamma^0 \sigma^2 \epsilon_0' = s \, \epsilon_0' \,, \qquad s =  \frac{\sin \alpha}{|\sin \alpha |} \,.
\end{equation}
This therefore correlates the sign of $q_m=p^Aq_A'-q_A p^A{}'$ with that of the projection applied to the constant spinor part of $\epsilon(x)$, whereas the sign of $q_e$---at fixed positive energy---determines the `side' of AdS$_2$ where the particle gets stabilized. We will slightly refine this statement in Section \ref{ss:susywithangularmomentum} below. Lastly, the combination of \eqref{eq:cond001} and \eqref{eq:cond003} yields the condition \cite{Simons:2004nm}
\begin{equation}
    \tanh{\chi} = \cos\alpha \,,
\end{equation}
which indeed coincides with eq.~\eqref{eq:dynamicalcondstatic}, since we set $\varphi=0$ and thus $\alpha$ defined in \eqref{eq:phasecentralcharge} measures the relative complex phase between the particle and black hole central charges.

\section{Supersymmetric Stationary Probes}\label{s:stationaryprobes}

In Section \ref{s:review}, we demonstrated that charged geodesics exhibiting non-trivial motion along $\mathbf{S}^2$ may be time-independent---in global AdS coordinates---if placed at certain radial positions in Anti-de Sitter space. This parallels the situation encountered when restricting ourselves to fully static configurations, where the particle remains still at some location on the sphere. Therefore, given that the latter trajectories preserve four out of the eight total supercharges of the background spacetime (cf. Section \ref{s:staticprobes}), it is natural to wonder whether more general stationary paths could also be supersymmetric. Our aim in this section will be to prove the latter statement. Additionally, we argue that the quantity determining which supersymmetries remain unbroken corresponds to the generalized angular momentum $\boldsymbol{J}$ of the system, thereby explaining when (and why) multi-particle states are mutually BPS. 

\subsection{Including orbital angular momentum}\label{ss:susywithangularmomentum}

To show that configurations with non-vanishing angular momentum can be supersymmetric, we proceed as in Section \ref{ss:recoveringStrominger}. Specifically, we seek to determine the dynamical conditions under which these trajectories break only half of the supersymmetries of the ambient space. The paths considered herein satisfy $\dot{\theta} = \dot{\chi}  = 0$ as well as $\dot{\tau}=1,\,\dot{\phi}=\pm1$ (cf. Section \ref{ss:geodesics}),\footnote{The sense of rotation is fixed by the direction of the generalized angular momentum, namely $\dot{\phi}=\text{sgn}(j)$.} such that the appropriate $\kappa$-symmetry projector is
\begin{equation}\label{eq:angmomentprojector}
    \Gamma_\kappa= \frac{1}{\sqrt{-h_{\tau \tau}}} \left( e_\tau{}^0\, \gamma_0 \pm e_\phi{}^3\, \gamma_3\right)=\frac{1}{\sqrt{\cosh^2\chi-\sin^2\theta}}\,\left(\cosh \chi\gamma_0 \pm\sin\theta\,\gamma_3\right)\, ,
\end{equation}
where we substituted both $\sqrt{-h_{\tau \tau}}= R\, (\cosh^2\chi-\sin^2 \theta)^{\frac12}$ and eq.~\eqref{eq:vierbeins}. On the other hand, as explained in Section \ref{ss:susy&kappasym}, the criterion \eqref{eq:kappaMaioranaImaginaryOriginal} for a classical trajectory to be BPS reads
\begin{equation}\label{eq:kappaMaioranaImaginary}
     \frac{1}{\sqrt{-h_{\tau \tau}}}\, \dot{x}^\mu e_\mu{}^a \, v^{-1}\,e^{- i \alpha \gamma_5}\,\gamma_a\, v\, \sigma^2  \epsilon_0= \epsilon_0\, ,
\end{equation}
where we used $\epsilon = v(x) \, \epsilon_0$ with $\epsilon_0$ a constant Majorana doublet and the spacetime dependent part being captured by $v(x)$. Comparing with \eqref{eq:KillingSpinorfinal}, we have 
\begin{equation}
    v = \Gamma_s(u \tilde{u})^{-1} =  e^{-\frac{1}{2}  \chi \gamma^0 \sigma^2} e^{\frac{1}{2} \tau \gamma^1 \sigma^2} e^{- \frac{1}{2}(\theta - \pi/2)  \gamma^0 \gamma^1 \gamma^2 \sigma^2} e^{-  \frac{1}{2 } \phi \gamma^0 \gamma^1 \gamma^3 \sigma^2 }\,.
\end{equation}
Thus, the relevant quantities needed to evaluate \eqref{eq:kappaMaioranaImaginary} are the following
\begin{equation}\label{eq:dressedmatrices}
    v^{-1} \, e^{-i \alpha \gamma_5} \gamma^0 \, v \, \sigma^2 \,, \qquad v^{-1} e^{-i \alpha \gamma_5} \gamma^3 \, v \, \sigma^2  \,,
\end{equation}
which, after some algebra (see Appendix \ref{app:identitiesABCD}), are shown to yield
\begin{equation}\label{eq:gamma0}
    \begin{split}
         v^{-1} e^{-i \alpha \gamma_5} \gamma^0 \, v \, \sigma^2 = & \; \cos \tau \left[ \, \cos \alpha \, \gamma^0 \sigma^2 - i \sin \alpha \sinh \chi \cos\theta \,  \gamma_5  \gamma^2 \gamma^1 \gamma^0  \sigma^2 \right] \\[2mm]
         &  + \sin \tau \left[ - \cos \alpha \, \gamma^1 \gamma^0  - i \sin \alpha \sinh \chi \cos\theta \,  \gamma_5  \gamma^2 \gamma^0 \right] \\[2mm]
         & + \cos\phi \left[- i \sin\alpha \cosh \chi \sin \theta \, \gamma_5 \gamma^0 \sigma^2 \right] \\[2mm]
         & + \sin \phi \left[- i \sin \alpha \cosh \chi \sin \theta \, \gamma_5 \gamma^3 \gamma^1   \right] \\[2mm]
         & + \cos \tau \cos \phi \left[\, i \sin \alpha \sinh \chi \sin \theta \, \gamma_5  \right] \\[2mm]
         & + \cos \tau \sin \phi \left[ \, i \sin \alpha \sinh \chi \sin \theta \, \gamma_5 \gamma^3 \gamma^1 \gamma^0 \sigma^2 \right] \\[2mm]
         & + \sin \tau \cos \phi \left[ \, i \sin \alpha \sinh \chi \sin \theta \, \gamma_5 \gamma^1 \sigma^2 \right] \\[2mm]
         & + \sin \tau \sin \phi \left[\, i \sin \alpha \sinh \chi \sin \theta \, \gamma_5 \gamma^3 \gamma^0   \right] \\[2mm]
         & + i \sin \alpha \cosh \chi \cos \theta \, \gamma_5 \gamma^2 \gamma^1 \,,
    \end{split}
\end{equation}
and
\begin{equation}\label{eq:gamma3}
    \begin{split}
         v^{-1} e^{-i \alpha \gamma_5} \gamma^3 \, v\, \sigma^2 = & \; \cos \tau \left[\, \cos \alpha \cosh \chi \sin \theta \, \gamma^3 \sigma^2 \right] + \sin \tau \left[\,\cos\alpha \cosh \chi \sin \theta \, \gamma^3 \gamma^1 \right] \\[2mm]
         & + \cos\phi \left[\,\cos \alpha \sinh \chi \cos \theta \, \gamma^3 \gamma^2 \gamma^1 \sigma^2 - i \sin \alpha \, \gamma_5 \gamma^3 \sigma^2 \right] \\[2mm]
         & + \sin \phi \left[ \, \cos \alpha \sinh\chi \cos \theta \,  \gamma^2 \gamma^0  + i \sin \alpha \, \gamma_5 \gamma^1 \gamma^0 \right] \\[2mm]
         & + \cos \tau \cos \phi \left[-\cos \alpha \cosh \chi \cos \theta \, \gamma^3 \gamma^2 \gamma^1 \gamma^0 \right] \\[2mm]
         & + \cos \tau \sin \phi \left[ - \cos \alpha \cosh \chi \cos \theta \, \gamma^2 \sigma^2 \right] \\[2mm]
         & + \sin\tau \cos \phi \left[ -\cos \alpha \cosh \chi \cos \theta \, \gamma^3 \gamma^2 \gamma^0 \sigma^2 \right] \\[2mm]
         & + \sin \tau \sin \phi \left[-\cos \alpha \cosh \chi \cos \theta \,\gamma^2 \gamma^1 \right] \\[2mm]
         & - \cos \alpha \sinh \chi \sin \theta \, \gamma^3 \gamma^0 \,.
    \end{split}
\end{equation}
To verify that eq.~\eqref{eq:kappaMaioranaImaginary} can be satisfied, we proceed by examining terms of different order in $\{\cos\tau, \sin\tau, \cos \phi, \sin \phi\}$ separately. From eqs.~\eqref{eq:gamma0} and \eqref{eq:gamma3}, one readily sees that both types of quadratic contributions, namely the ones picking up a sign under $(\tau, \phi) \to (-\tau,-\phi)$ as well as those which are left invariant, lead to the exact same condition. For instance, the invariant terms under the $\mathbb{Z}_2$-map give rise to a piece in the l.h.s. of \eqref{eq:kappaMaioranaImaginary} of the form
\begin{equation}
\begin{aligned}
    &\frac{R\cos \tau \cos \phi}{\sqrt{-h_{\tau \tau}}}\left(-i \sin \alpha \cosh\chi \sinh\chi\sin\theta \mp i\cos\alpha \cosh \chi \cos\theta \sin\theta\right)\gamma_5 \epsilon_0\\
    &+\frac{R\sin \tau \sin \phi}{\sqrt{-h_{\tau \tau}}}\left(-i \sin \alpha \cosh\chi \sinh\chi\sin\theta \mp i\cos\alpha \cosh \chi \cos\theta \sin\theta\right)\gamma_5 \gamma^3 \gamma^0\epsilon_0\, ,
\end{aligned}
\end{equation}
which must hence vanish identically. This requires to have
\begin{equation}\label{eq:dyncondkappasymm}
    \tan\alpha\, \sinh\chi= \mp \cos\theta\,.
\end{equation}
Note that this resembles the relation obtained in the static case (cf. eq.~\eqref{eq:cond001}), which may be embedded within the stationary configuration above by placing the particle at one of the poles of $\mathbf{S}^2$, see Figure \ref{sfig:static}. In fact, \eqref{eq:dyncondkappasymm} is seen to exactly reproduce the latter when taking the limit of zero orbital angular momentum, i.e., $\ell \to 0$. 

Subsequently, and inspired by the static case, we impose the following restriction on the constant part of the Killing spinor (cf. eq.~\eqref{eq:projectorstatic})\footnote{The precise form of the projection operator acting on $\epsilon_0$ is determined in general by the direction of the generalized angular momentum $\boldsymbol{J}$, and can be obtained from \eqref{eq:projstationary} by applying an appropriate $SU(2)$ rotation, see discussion around eq.~\eqref{eq:projectorstatic}.}
\begin{equation}\label{eq:projstationary}
    \gamma^3 \gamma^0 \epsilon_0=\mp \epsilon_0 \, .
\end{equation}
It should be stressed that the signs in the two previous equations are necessarily correlated to each other so as to ensure that \eqref{eq:kappaMaioranaImaginary} is ultimately satisfied. This becomes particularly clear when considering terms which depend linearly on $\{\cos\tau, \sin\tau, \cos \phi, \sin \phi\}$. However, before proceeding any further, let us remark that we can already understand at this point what physical quantity determines the precise projection on $\epsilon_0$ by substituting the stationary geodesics of Section \ref{sss:static&stationarypaths}, which are such that 
\begin{equation}\label{eq:dynamicalcondstationary}
    \tan\alpha=\frac{q_m}{q_e}\, ,\qquad \sinh\chi= \frac{q_e}{\sqrt{j^2}}\, ,\qquad \cos\theta =-\frac{q_m}{j}\, ,
\end{equation}
into eq.~\eqref{eq:dyncondkappasymm} above. Therefore, denoting by $s$ the sign to be chosen in \eqref{eq:projstationary}, one obtains
\begin{equation}\label{eq:signprojectorstationary}
    s= -\frac{j}{\sqrt{j^2}}=-\,\text{sgn}(j)\, .
\end{equation}
In particular, this implies that the supersymmetries that are left unbroken are determined by the direction of the generalized angular momentum associated to the particle. Interestingly, this accommodates the static case discussed previously where, by placing the particle in one of the two poles of the sphere and using eqs.~\eqref{eq:projectorstatic} and \eqref{eq:signprojectorstatic}, we are lead to
\begin{equation}
    s= \text{sgn}(q_m \cos \theta)= -\, \text{sgn}(j)\, ,
\end{equation}
in agreement with \eqref{eq:signprojectorstationary} above. This observation clarifies why particles and antiparticles, whose electric and magnetic charges are equal in magnitude but opposite in sign, preserve identical supersymmetries when located at antipodal points on $\mathbf{S}^2$. The resulting configurations possess the same generalized angular momentum! Consequently, in the more general scenario where the D-brane probes also move on the sphere, one may obtain BPS multi-particle states \textit{if and only if} the absolute value of the total angular momentum $\boldsymbol{J}$ of the system equals the sum of the angular momenta of the individual constituents 
\begin{equation}
    |\boldsymbol{J}_{\rm tot}|=\sum_i |\boldsymbol{J}_{i}|\, ,
\end{equation}
since they all project out the same set of supercharges, see Figure \ref{sfig:stationary}.

\medskip

In the remainder, we demonstrate that condition \eqref{eq:dyncondkappasymm} together with \eqref{eq:projstationary} suffice to ensure that eq.~\eqref{eq:kappaMaioranaImaginary} holds. Indeed, focusing on the linear terms in $\{\cos\tau, \sin\tau, \cos \phi, \sin \phi\}$ we find
\begin{equation}
\begin{aligned}
    &\frac{R\cos \tau}{\sqrt{-h_{\tau \tau}}}\left(\mp \cos \alpha \cosh\chi -\sin\alpha \cosh\chi \sinh\chi\cos\theta \pm \cos\alpha \cosh \chi \sin^2\theta\right)\gamma^3 \sigma^2 \epsilon_0\\
    &+\frac{R\cos \phi}{\sqrt{-h_{\tau \tau}}}\left(\pm i \sin \alpha \cosh^2\chi \sin\theta + i\cos\alpha \sinh \chi \cos\theta \sin\theta \mp i\sin\alpha \sin\theta\right)\gamma_5 \gamma^3 \sigma^2\epsilon_0\, , 
\end{aligned}
\end{equation}
as well as
\begin{equation}
\begin{aligned}
    &\frac{R\sin \tau}{\sqrt{-h_{\tau \tau}}}\left( \cos \alpha \cosh\chi \pm \sin\alpha \cosh\chi \sinh\chi\cos\theta - \cos\alpha \cosh \chi \sin^2\theta\right)\gamma^1 \gamma^0 \epsilon_0\\
    &+ \frac{R\sin \phi}{\sqrt{-h_{\tau \tau}}}\left( \sin \alpha \cosh^2\chi \sin\theta \pm \cos\alpha \sinh \chi \cos\theta \sin\theta - \sin\alpha \sin\theta\right)\gamma^2 \gamma^0 \epsilon_0\, ,
\end{aligned}
\end{equation}
which trivially cancel. Lastly, the time-independent component would read as follows
\begin{equation}
\begin{aligned}
    &\frac{R}{\sqrt{-h_{\tau \tau}}}\left( \sin \alpha \cosh^2\chi \cos\theta \mp\cos\alpha \sinh\chi \sin^2\theta\right)\gamma^3 \gamma^0 \epsilon_0\, ,
\end{aligned}
\end{equation}
such that inserting eq.~\eqref{eq:dyncondkappasymm} and using that $h_{\tau\tau}=-R^2\sinh^2 \chi \sec^2 \alpha$, we finally arrive at
\begin{equation}
    \gamma^3 \gamma^0 \epsilon_0= \mp \epsilon_0 \,,
\end{equation}
being this again verified as per \eqref{eq:projstationary}.

\medskip

This concludes our proof that the classical stationary paths introduced in Section \ref{sss:static&stationarypaths} preserve half of the superconformal symmetries of AdS$_2\times \mathbf{S}^2$ solutions. In the next subsection, we will show that the trajectories satisfying eq.~\eqref{eq:dyncondkappasymm} also saturate a lower bound for the worldline Hamiltonian, implying that they can equivalently be determined---up to $SU(2)$ rotations---by minimizing the global energy of the particle probe.

\subsection{Saturating a BPS bound}\label{ss:BPSbound}

Let us consider the wordline action of a dyonic particle in 4d $\mathcal{N}=2$ with mass $m$ and gauge charges $(p^A, q_A)$. Specifying the latter to the near-horizon geometry of a BPS black hole described in global coordinates one obtains \eqref{eq:globalwordlineactionAdS2xS2}. If we also restrict to the static gauge, where we use the global time as worldline parameter---namely $\sigma=\tau$, the latter reduces to 
\begin{equation}
 S_{wl} = - \int d\tau \left[ \tilde{m}\,\sqrt{\cosh^{2}\chi-\left(\frac{d \chi}{d\tau}\right)^2 -\left(\frac{d\theta}{d\tau}\right)^2-\sin^2\theta \left(\frac{d \phi}{d\tau}\right)^2} - q_e\sinh\chi + q_m \cos \theta\, \frac{d \phi}{d\tau}\right]\, .
\end{equation}
From here one may readily compute the Hamilton operator
\begin{equation}\label{eq:globaltimeHamiltonian}
 \mathcal{H} = \mathsf{P}_i\, \frac{d x^i}{d\tau} - \mathcal{L} =\cosh \chi\sqrt{\tilde{m}^2+  \mathsf{P}_\chi^2+ \mathsf{P}_\theta^2 + \csc^2 \theta \left( \mathsf{P}_\phi + q_m \cos \theta\right)^2} -q_e\sinh\chi\, ,
\end{equation}
with the conjugate momenta being
\begin{equation}
 \mathsf{P}_\chi= \frac{\tilde{m}}{\sqrt{-h_{\tau \tau}}} \frac{d \chi}{d\tau}\, ,\qquad \mathsf{P}_\theta= \frac{\tilde{m}}{\sqrt{-h_{\tau \tau}}} \frac{d \theta}{d\tau}\, ,\qquad \mathsf{P}_\phi= \frac{\tilde{m}}{\sqrt{-h_{\tau \tau}}} \sin^2 \theta \frac{d \phi}{d\tau} - q_m \cos \theta\, ,
\end{equation}
whilst 
\begin{equation}
\begin{aligned}
 h_{\tau \tau}=g_{\mu \nu}\frac{d x^\mu}{d\tau} \frac{d x^\nu}{d\tau} = \frac{-\tilde{m}^2 \cosh^2 \chi}{\tilde{m}^2+\mathsf{P}_\chi^2+ \mathsf{P}_\theta^2 + \csc^2 \theta \left( \mathsf{P}_\phi + q_m \cos \theta\right)^2}\, ,
\end{aligned}
\end{equation}
denotes the pull-back of the spacetime metric onto the worldline. Furthermore, using the explicit form of the conserved angular momentum along the 2-sphere \cite{Castellano:2025yur}
\begin{equation}
\begin{aligned}
    &\mathsf{J}_1 = -\sin \phi\, \mathsf{P}_\theta-\cot \theta \cos\phi\, \mathsf{P}_\phi - q_m \csc \theta \cos \phi\, ,\\
    &\mathsf{J}_2 = \cos \phi\, \mathsf{P}_\theta-\cot \theta \sin\phi\, \mathsf{P}_\phi - q_m \csc \theta \sin \phi\, ,\\
    &\mathsf{J}_3 = \mathsf{P}_\phi\, ,
    \end{aligned}
\end{equation}
the Hamiltonian \eqref{eq:globaltimeHamiltonian} can be written as follows
\begin{equation}
 \mathcal{H} =\cosh \chi\sqrt{\tilde{m}^2-q_m^2+ \mathsf{P}_\chi^2+ \mathsf{J}^2} - q_e\sinh\chi\, .
\end{equation}
Notice that the minimum value for $\mathcal{H}$ occurs when $\mathsf{P}_\chi=0$ and $\tanh\chi= q_e/\sqrt{q_e^2 + \mathsf{J}^2}$, where we have imposed the BPS condition $\tilde{m}^2=q_e^2+q_m^2$, namely for stationary solutions (in AdS$_2$) satisfying \eqref{eq:radialequilibriumangmom}. Indeed, we find that for those trajectories the Hamiltonian saturates the following BPS bound
\begin{equation}
 \mathcal{H} \geq \sqrt{\mathsf{J}^2}\, .
\end{equation}
Classically, this defines a continuum of supersymmetric states labeled by the quadratic Casimir on the sphere. Quantum mechanically, though, the possible (generalized) angular momenta get quantized, defining different selection sectors that should have an analogue in the BPS spectrum of the putative dual CFT$_1$.\footnote{In those cases where a CFT$_2$ dual exists \cite{Maldacena:1997de}, the stationary BPS configurations of interest can be identified with chiral primary states in the CFT having non-vanishing $SU(2)$ R-charge \cite{Gaiotto:2006ns}.}

\section*{Acknowledgements}
	
We acknowledge valuable conversations with José Calderón-Infante, Damian van de Heisteeg, Juan Maldacena, Tomás Ortín, Sav Sethi, and Max Wiesner. A.C. thanks the Aspen Center for Physics, funded by the NSF grant PHY-2210452, for hospitality during the last stages of this work. The work of A.C. is supported by a Kadanoff and an Associate KICP fellowships, as well as through the NSF grants PHY-2014195 and PHY-2412985. A.C. and M.Z. are also grateful to Teresa Lobo and Miriam Gori for their continuous encouragement and support.
	
	
\appendix

\section{Conventions on 4d Spinors}\label{ap:conventions}	

In this work, we employ the mostly plus signature $(-, +,+,+)$ for the metric tensor in $d=4$. We also adopt the Majorana representation for the gamma matrices and, in particular, choose them to be purely imaginary. They thus satisfy the Clifford algebra
\begin{equation}
    \{\gamma^a, \gamma^b \} = - 2 \eta^{ab} \,,
\end{equation}
with $\eta$ the (mostly plus) Minkowski metric. In our conventions, the Dirac matrices verify
\begin{equation}\label{eq:unitaryrep}
    (\gamma^0)^2 = -(\gamma^i)^2 =  \mathbf{1}_4 \,, \qquad (\gamma^a)^* = - \gamma^a \,, \qquad (\gamma^0)^{\dagger}  = \gamma^0 \,, \qquad (\gamma^i)^{\dagger} = - \gamma^i \,,
\end{equation}
where $i = 1,2,3$, and we split the indices as $a = (0,i)$. In addition, we define the fifth gamma matrix
\begin{equation}
    \gamma_5 = - i \gamma^0 \gamma^1 \gamma^2 \gamma^3 \,, \qquad \qquad \text{such that}\quad (\gamma_5)^2 =  \mathbf{1}_4 \,,
\end{equation}
as well as the totally antisymmetric product
\begin{equation}
    \gamma^{a_1 a_2 \dots a_n} = \gamma^{[a_1} \gamma^{a_2} \dots \gamma^{a_n]} \,. 
\end{equation}
Notice that, for $ 1 \le n \le 4 $, one can prove the following identity 
\begin{equation}\label{eq:AppRel1}
    \gamma^{a_1 \dots a_n} = \frac{i}{(4-n)!}(-1)^{\left\lfloor \frac{n-1}{2} \right\rfloor} \epsilon^{a_1 \dots a_n }{}_{b_1 \dots b_{4-n}} \gamma^{b_1 \dots b_{4-n}} \gamma_5 \,,
\end{equation}
where $\left\lfloor \cdot \right\rfloor$ is the integer part function, and the normalization of the Levi-Civita symbol is
\begin{equation}\label{eq:orientation}
    \epsilon^{0123} = -1 \,.
\end{equation}
In the particular case of $n = 3$, we also have the useful relation
\begin{equation}\label{eq:AppRel2}
    \gamma^{abc} = \gamma^a \gamma^b \gamma^c + \gamma^a \eta^{bc} - \gamma^b \eta^{ac} + \gamma^c \eta^{ab} \,.
\end{equation}
The charge conjugation matrix $\mathcal{C}$ we use to impose the Majorana condition on spinors is\footnote{This matrix is used to define the charge conjugate $\psi^c= \mathcal{C} \bar{\psi}^T$, and in the representation \eqref{eq:unitaryrep} it verifies 
\begin{equation}
    \mathcal{C}^\dagger \mathcal{C} = 1\,,\qquad \mathcal{C} \gamma_\mu\mathcal{C}^{-1}=-\gamma_\mu^T\, .\notag
\end{equation}}
\begin{equation}
    \mathcal{C} = i \gamma_0 \,.
\end{equation}
On the other hand, the spinor doublets in the Weyl representation satisfy
\begin{equation}
    \epsilon_A = (\epsilon^A )^* \,, \qquad \gamma_5 \, \epsilon^A = \epsilon^A \,, \qquad \gamma_5 \epsilon_A = -\epsilon_A \,.
\end{equation}
They are related to the Majorana fermions $\epsilon^i$ via the map
\begin{equation} \label{eq:AppMapToMajorana}
    \epsilon^i = \epsilon^I \oplus \epsilon_I \,, \qquad i, I = 1,2 \,,
\end{equation}
whose inverse relation reads instead
\begin{equation}
    \epsilon_I = \frac{1}{2}\left(1 - \gamma_5 \right) \epsilon^i \,, \qquad  \epsilon^I = \frac{1}{2}\left(1 + \gamma_5 \right)\epsilon^i  \,.
\end{equation}
It is useful to recall how one can determine whether we can further constrain a Majorana spinor using a projection operator. Hence, suppose we have an involution matrix such that $\Omega^2 = 1$ and we want to impose a condition on $\epsilon$ of the form $\Omega \, \epsilon = \pm \epsilon$. The projection with $P_\pm =\frac12 (1\pm\Omega)$ is compatible with the reality of the spinor provided that $\Omega$ verifies the relation 
\begin{equation} \label{eq:compatibilityCondition}
    \mathcal{C}^{-1} \Omega^T \mathcal{C} = \gamma_0 \Omega^\dagger \gamma_0 \,.
\end{equation}
Notice, for instance, that if we pick $\Omega = \gamma_5$ then condition \eqref{eq:compatibilityCondition} is not satisfied. And indeed in four-dimensional, Lorentzian spacetime we can not have Weyl-Majorana spinors.

\section{Useful Identities Involving Dirac Matrices}\label{app:identitiesABCD}

The aim of this appendix is to list several mathematical identities and formal manipulations concerning the Clifford algebra (see Appendix \ref{ap:conventions} for our conventions) that become useful when performing the computations outlined in sections \ref{s:staticprobes} and \ref{s:stationaryprobes}.

\medskip

Therefore, consider the spinorial matrices
\begin{align}
    A = e^{-\frac{1}{2} \chi \gamma^0 \sigma^2} \,, \qquad    
    B = e^{\frac{1}{2} \tau \gamma^1 \sigma^2}  \,, \qquad C = e^{-\frac{1}{2} (\theta - \frac{\pi}{2}) \gamma^0 \gamma^1 \gamma^2 \sigma^2} \,, \qquad D = e^{-\frac{1}{2} \phi \gamma^0 \gamma^1 \gamma^3 \sigma^2} \,.
\end{align}
which encode the spacetime dependence of the Killing spinors in AdS$_2\times \mathbf{S}^2$, cf. eq.~\eqref{eq:KillingSpinorfinal}. They satisfy the following (anti)commutation relations
\begin{equation}
\begin{aligned}
[\gamma^0, A] &= 0,           &\qquad \gamma^i A &= A^{-1} \gamma^i, &\qquad i &= 1,2,3\,, \\[2mm]
[\gamma^1, B] &= 0,           &\qquad \gamma^i B &= B^{-1} \gamma^i, &\qquad i &= 0,2,3\,, \\[2mm]
[\gamma^i, C] &= 0,           &\qquad \gamma^3 C &= C^{-1} \gamma^3, &\qquad i &= 0,1,2\,, \\[2mm]
[\gamma^i, D] &= 0,           &\qquad \gamma^2 D &= D^{-1} \gamma^2, &\qquad i &= 0,1,3\,,
\end{aligned}
\label{eq:gamma_relations}
\end{equation}
and 
\begin{equation}
\begin{aligned}
[\gamma_5 \gamma^i, A] &= 0,           &\qquad \gamma_5 \gamma^0 A &= A^{-1} \gamma_5 \gamma^0, &\qquad i &= 1, 2,3\,, \\[2mm]
[\gamma_5 \gamma^i, B] &= 0,           &\qquad \gamma_5 \gamma^1 B &= B^{-1} \gamma_5 \gamma^1, &\qquad i &= 0,2,3\,, \\[2mm]
[\gamma_5 \gamma^3, C] &= 0,           &\qquad \gamma_5 \gamma^i C &= C^{-1} \gamma_5 \gamma^i, &\qquad i &= 0,1,2\,, \\[2mm]
[\gamma_5 \gamma^2, D] &= 0,           &\qquad \gamma_5 \gamma^i D &= D^{-1} \gamma_5 \gamma^i, &\qquad i &= 0,1,3\,.
\end{aligned}
\label{eq:gamma5_relations}
\end{equation}
Notice that the above relations imply that $A,B$ commute with $C,D$ (equiv. their conjugates). Their squares can also be written more compactly as follows
\begin{subequations}
\begin{align}
    A^{-2} & = e^{\chi \gamma^0 \sigma^2} = \cosh \chi + \sinh \chi \gamma^0 \sigma^2 \,, \\[2mm]
    B^{-2} & = e^{-\tau \gamma^1 \sigma^2} = \cos \tau - \sin \tau \gamma^1 \sigma^2  \,, \\[2mm]
    {C}^{-2} & = e^{(\theta - \pi/2) \gamma^0 \gamma^1 \gamma^2 \sigma^2} = \sin\theta - \cos \theta\, \gamma^0 \gamma^1 \gamma^2 \sigma^2 \,, \\[2mm]
    {D}^{-2} & = e^{\phi \gamma^0 \gamma^1 \gamma^3 \sigma^2} = \cos \phi + \sin \phi \gamma^0 \gamma^1 \gamma^3 \sigma^2 \,.
\end{align}
\end{subequations}
Finally, another useful relation that is thoroughly used throughout the main text is
\begin{equation}
    e^{-i \alpha \gamma_5} = \cos\alpha - i \sin \alpha \gamma_5 \,,
\end{equation}
where $\alpha$ frequently denotes the phase of the central charge of the particle, see e.g., eq.~\eqref{eq:phasecentralcharge}.

\bibliography{ref.bib}
\bibliographystyle{JHEP}

\end{document}